\RequirePackage{ifpdf}
\documentclass{JHEP3}
\usepackage{amsfonts,amsbsy,latexsym,amssymb,amscd,amstext}
\usepackage{epsfig}
\usepackage{graphicx}
\usepackage{amsmath}
\usepackage{bm}
\usepackage{placeins}
\usepackage{verbatim}   
\usepackage{cleveref}

\newcommand{\HG}{\hat{\Gamma}}
\newcommand{\TGF}{{\rm TGF}}
\newcommand{\be}{\begin{equation}}
\newcommand{\ee}{\end{equation}}
\newcommand{\ba}{\begin{array}}
\newcommand{\ea}{\end{array}}
\newcommand{\baa}{\begin{array}}
\newcommand{\eaa}{\end{array}}
\newcommand{\bea}{\begin{eqnarray}}
\newcommand{\eea}{\end{eqnarray}}

\newcommand{\Tr} {{\rm Tr}}

\newcommand{\lt}{{\tilde l}}

\title{The $SU(\infty)$ twisted gradient flow running coupling}

\author{ Margarita Garc\'{\i}a P\'erez$^{a}$, Antonio Gonz\'alez-Arroyo$^{a,b}$,
Liam Keegan$^{c}$ \quad \quad \quad \quad \quad \quad and Masanori Okawa$^{d}$ \\
  $^a$ Instituto de F\'{\i}sica Te\'orica UAM-CSIC, Nicol\'as Cabrera 13-15, \\
  Universidad Aut\'onoma de Madrid, E-28049--Madrid, Spain \\
  $^b$ Departamento de F\'{\i}sica Te\'orica, C-XI \\
       Universidad Aut\'onoma de Madrid, E-28049--Madrid, Spain \\
  $^c$ PH-TH, CERN, CH-1211 Geneva 23, Switzerland\\
  $^d$ Graduate School of Science, Hiroshima University,\\
Higashi-Hiroshima, Hiroshima 739-8526, Japan \\

E-mail: \email{margarita.garcia@uam.es,antonio.gonzalez-arroyo@uam.es,liam.keegan@cern.ch, okawa@sci.hiroshima-u.ac.jp}
 }

\abstract{We measure the running of the $SU(\infty)$ 't Hooft coupling by performing a step scaling 
analysis of the Twisted Eguchi-Kawai (TEK) model, the SU($N$) gauge theory on a 
single site lattice with twisted boundary conditions. 
The computation relies on the conjecture that finite volume effects for SU(N) gauge theories defined
on a 4-dimensional twisted torus are controlled by an effective size parameter $\tilde l = l \sqrt{N}$, with $l$ the torus period.
We set the scale for the running coupling in terms of $\tilde l$ 
and use the gradient flow to define a renormalized 't Hooft coupling $\lambda(\lt)$.
In the TEK model, this idea allows the determination of the running of the coupling through a step scaling procedure that uses the rank of the group as a size parameter.
The continuum renormalized coupling constant is extracted in the zero lattice spacing limit, which in the TEK model corresponds to the large $N$ 
limit taken at fixed value of $\lambda(\lt)$. The coupling constant is thus expected to coincide with that of the ordinary pure gauge theory at $N =\infty$.
The idea is shown to work and permits us to follow the evolution of the coupling over a wide range of scales.
At weak coupling we find a remarkable agreement with the perturbative two-loop formula for the running coupling.}


\preprint{CERN-PH-TH-2014-233\\
IFT-UAM/CSIC-14-125\\ FTUAM-14-50 \\HUPD-1411}

\date{\today}

\begin{document}

\section{Introduction}

The Twisted Eguchi-Kawai (TEK) model \cite{GonzalezArroyo:1982ub,GonzalezArroyo:1982hz,GonzalezArroyo:2010ss} is a single-site formulation of SU($N$) 
lattice gauge theory. In the large $N$ limit, taken at fixed bare 't Hooft coupling, it becomes equivalent to 
a $SU(\infty)$ lattice gauge theory in the thermodynamic limit, as tested numerically in detail in 
Ref. \cite{Gonzalez-Arroyo:2014dua}. In this paper we will be concerned with the analysis of 
the TEK model in a different scaling regime. 
The conjecture of TEK volume reduction and the more general one of volume independence 
at finite $N$ with twisted boundary conditions have been recently reviewed in Ref. \cite{Perez:2014sqa}, and analyzed 
in 2+1 dimensions in Refs. \cite{Perez:2013dra,Perez:2012fz}. The main ingredient to be used in this paper is that
SU(N) gauge theories, when defined on twisted 4-dimensional tori, have volume effects
controlled by an effective size parameter $\tilde l = l \sqrt{N}$, with $l$ the torus period. 
Our objective will be to determine the non-perturbative running of the 't Hooft coupling with the effective 
scale $\tilde l$. For that purpose we will be using a standard step scaling procedure $\lt \rightarrow s \lt$ 
implemented non-perturbatively by discretizing the torus on a lattice \cite{Luscher:1991wu}. 
The unusual feature in our determination is that we will work on a single-site TEK lattice
with $\tilde l = a \sqrt{N}$. Even in this extreme case, volume independence suggests that step scaling may
be implemented by scaling the gauge group from $SU(N)$ to $SU(s^2N)$. We will show this procedure at work and 
will reproduce the 2-loop running of the coupling constant from the step scaling
non-perturbative simulations. One important 
remark is that this will require us to approach the continuum limit at a fixed value of $\lt$, which amounts in the TEK model to a large $N$ limit taken
at fixed renormalized 't Hooft coupling $\lambda(\lt)$. If volume independence holds, we expect that our calculation will 
provide the running of the $SU(\infty)$ 't Hooft coupling in the continuum limit.

Before proceeding any further, let us mention that twisted boundary conditions have already been used 
in combination with the Yang-Mills gradient flow \cite{Narayanan:2006rf,Luscher:2010iy}
to define a running coupling for $SU(N)$ gauge theories \cite{Ramos:2013gda,Ramos:2014kla}. 
Here we define an analogous coupling that runs in terms of the effective scale $\tilde l$.  
Preliminary results of this work have been presented in Ref. \cite{Perez:2014xga}.

The paper is organized as follows. Sec. \ref{s.vol} discusses briefly the idea of volume independence, linking finite size and finite $N$ effects in the
presence of twisted boundary conditions. We discuss how this general idea particularizes to the case of the TEK model.  In sec. \ref{s.flow} we define 
a non-perturbative coupling based on the use of the gradient flow on a 4-dimensional torus with twisted boundary conditions in all directions. The 
renormalization scale is fixed in terms on the effective box size $\lt$.
The perturbative behaviour of the gradient flow in this set-up is analyzed in sec. \ref{s.disc}. We briefly discuss how to improve the lattice definition of the 
coupling at tree-level in perturbation theory deferring all the technical details to Appendix \ref{ap.a}. 
In sec. \ref{s.results} we present the results of a non-perturbative calculation of the TEK running coupling and describe in detail the step scaling procedure
involved in its determination. We conclude with a brief summary of our results. Appendix \ref{ap.b} collects all our numerical data.

\section{Volume independence in $SU(N)$ gauge theories on the twisted torus}
\label{s.vol}

Twisted Eguchi-Kawai reduction \cite{GonzalezArroyo:1982ub,GonzalezArroyo:1982hz,GonzalezArroyo:2010ss} can be considered a particular case of the more general idea of
volume independence in Yang-Mills theories with twisted boundary conditions, recently reviewed in Ref. \cite{Perez:2014sqa}. 
The main observation is that finite size and finite $N$ effects are intertwined. 
In the case of the 4-dimensional lattice TEK model, the corrections at finite $N$ take the form, in perturbation theory, of finite volume corrections 
for an effective lattice size of $\sqrt{N}$. For instance, the propagator is identical to that of a 
$(\sqrt{N})^4$ lattice \cite{GonzalezArroyo:1982hz}.  In this paper we 
will use this fact to define a running coupling constant in the large $N$ gauge theory using the rank of the group as a size parameter.

To be precise, let us start from the general case of a $SU(N)$ gauge theory defined on a four dimensional 
torus with all periods equal to $l$ and twisted boundary conditions.
It has been conjectured that finite size effects are controlled by an effective size parameter 
given by: $\lt = l \sqrt{N}$. This is so for the set of irreducible antisymmetric twist tensors \cite{GonzalezArroyo:1997uj}:
\be
\label{eq.twist}
n_{\mu\nu} =\, k  \sqrt{N} \, , \, \, {\rm for} \ \mu<\nu \, ,
\ee
with $k$ and $\sqrt N$ coprime integers. 
The conjecture is sustained by the observation that  
the momentum quantization rule and the free propagators correspond to those of a box with extended periods $\tilde l$.  
Moreover, the perturbative Feynman rules in the twisted box respect the $\lt$ dependence up 
to a phase factor determined by the boundary conditions through the angle:
\be 
\tilde \theta = {2\pi |\bar k| \over \sqrt N} \, ,
\ee
with $\bar k$ defined to satisfy $k \bar k = 1\, (\mathrm{mod}\,\, \sqrt N)$. 
Provided $\tilde \theta$ is kept fixed as the large $N$ limit is taken, volume effects in perturbation theory are
controlled by the effective size $\lt$. 

In order to establish a connection to the TEK model, one discretizes the theory on a 
$L^4$ lattice with $\lt= a L \sqrt{N}$. The TEK model corresponds 
to the case of a one point lattice with $L=1$. It is defined in terms of four SU(N) matrices $U_{\mu}$, with the action
\begin{equation}
\label{eq.tek}
 S = bN \sum_{\mu\nu}\left(N - Z_{\mu\nu} \Tr \left[U_{\mu}U_{\nu}U^{\dagger}_{\mu}U^{\dagger}_{\nu}\right] \right), \quad Z_{\mu\nu} = Z^{*}_{\nu\mu} = e^{2\pi i k/\sqrt{N}}\,\,\mathrm{for}\,\,\mu<\nu,
\end{equation}
where $b$ is the lattice analog of the inverse 't Hooft coupling, $ 1/(Ng^2)$. 
In the original proposal, put forward long ago in Ref. \cite{GonzalezArroyo:1982hz}, 
the large $N$ limit is attained at fixed value of the lattice spacing $a$, with the continuum limit taken afterwards
driven by the large $N$ beta function. TEK reduction implies that the resulting theory is equivalent to a $SU(\infty)$ gauge theory
in the thermodynamic limit. This holds as long as center symmetry is not spontaneously broken for large $N$, i.e. 
the trace of all open Wilson loops on the lattice should go to zero in this limit. For that to be the case the flux $k$ has to satisfy 
$k/\sqrt{N} > 1/9$ \cite{GonzalezArroyo:2010ss}.  
As mentioned in the introduction, we will follow a different strategy, taking the continuum limit at a fixed value of the
effective torus size $\tilde l$. For the  TEK model $\tilde l = a \sqrt{N}$, and the continuum limit  corresponds to the $N\rightarrow \infty $
limit taken at fixed $\tilde l$. This has to be done while scaling the flux appropriately 
to keep the parameter $\tilde{\theta}$ fixed \cite{Perez:2014sqa}.  

\section{Twisted Gradient Flow (TGF) running coupling: $\lambda_{\TGF}$}
\label{s.flow}

To determine the running coupling  we will make use of the recently proposed Twisted Gradient 
Flow (TGF) scheme \cite{Ramos:2014kla}. 
The gradient flow \cite{Narayanan:2006rf,Luscher:2010iy} smooths gauge fields along a flow-time trajectory 
defined by the equation:
\be
\partial_t B_\mu(x,t) = D_\nu G_{\nu\mu} (x,t)
\ee
with $B_\mu(x,t=0)$ determined by the gauge potential $A_\mu(x)$.
At positive gradient flow time, the action density of SU($N$) gauge theory 
is a renormalized quantity with a perturbative expansion in the thermodynamic limit given by \cite{Luscher:2010iy,Luscher:2011bx},
\be
\tfrac{1}{N} \left\langle E(t) \right\rangle=\tfrac{1}{2N} \left\langle \Tr\{G_{\mu\nu}(t)G_{\mu\nu}(t)\} \right\rangle 
= \frac{3(N^2-1)}{128 N^2 \pi^2 t^2} \, \lambda_{\overline{MS}} + \mathcal{O}(\lambda^2_{\overline{MS}}),
\ee
with $\lambda_{\overline{MS}}\equiv N g^2_{\overline{MS}}$ denoting the 't Hooft coupling in the $\overline{MS}$ scheme. 
This quantity can be used to define a renormalized coupling at a renormalization scale $\mu=1/\sqrt{8t}$. 
The identification of this scale with the linear size of the box gives rise to the finite volume gradient flow schemes
\cite{Fodor:2012td,Fritzsch:2013je}. In this context, the use of twisted boundary conditions, leading
to the TGF scheme,  has many advantages \cite{Ramos:2014kla}. Among them, the absence of zero momentum modes 
and the manifest invariance of the theory under space-time translations. In this paper we will present a 
modification of the TGF scheme which adopts the twisted boundary conditions introduced in the 
previous section. It incorporates the idea of
volume independence by fixing the renormalization scale in terms of the effective box size $\tilde l = l \sqrt{N}$.
The renormalized coupling at scale $\tilde l$ is thus given by:
\begin{equation}
\label{lambda}
\lambda_{\TGF}(\tilde l) = \left. \mathcal{N}^{-1}(c) \, \frac{t^2 \langle E (t) \rangle }{N} \right|_{t=c^2 
{\tilde l}^2 /8} \, ,
\end{equation}
with $c$ an arbitrary constant parameter defining the renormalization scheme.
The constant $\mathcal{N}(c)$ is determined by matching $\lambda_{\TGF} (\tilde l)$ to the bare 't Hooft  
coupling ($\lambda_0$) at tree-level in perturbation theory.
 The details of the calculation of $\mathcal{N}(c)$ on a finite twisted torus are 
presented in Appendix \ref{ap.a}. The tree-level expansion of $E(t)$ is easily obtained in momentum space:
\be
\frac{t^2 \langle E(t) \rangle}{N} \Big |_{\rm tree} =   {\frac{3 \lambda_0 \, t^2}{2 \, {\tilde l}^4}} \sum'_q  e^ {-2t q^2 }\, ,
\ee
where $q_\mu= 2\pi n_\mu /\tilde l$, with $n_\mu \in Z\!\!\! Z$.  The prime in the sum implies the exclusion of momenta with
$n_\mu = 0 \,  ({\rm mod} \  \sqrt{N})$, $\forall \mu$. This leads to:
\be
\mathcal{N}(c) =  \frac{ 3 c^4 }{128} \Big (\theta_3^4(0, i \pi c^2 )- \theta_3^4(0, i \pi c^2 N ) \Big)\, ,
\ee
expressed in terms of the Jacobi Theta function:
\be
\theta_3(z,it) =  \sum_{m\in Z\!\!\!Z} e^ {- t \pi m^2 + 2 \pi i m z}\, .
\ee

A non-perturbative determination of the running coupling requires a lattice calculation. Our proposal is to replace
the standard step scaling procedure \cite{Luscher:1991wu} taking into account that the effective box size is $\lt$. Accordingly 
we define a continuum step scaling function
\be
\sigma(u,s) = \left.\lambda_{\rm TGF}(s\lt)\right|_{\lambda_{\rm TGF}(\lt)=u}\, ,
\ee
and the corresponding lattice expression
\be
\Sigma(u,s,L\sqrt{N}) = \left.\lambda_{\rm TGF}(sL\sqrt{N},b)\right|_{\lambda_{\rm TGF}(L\sqrt{N},b)=u}
\ee
defined on an $L^4$ site lattice with $\tilde l = a L \sqrt{N}$.
In addition, we will push the idea of volume independence to the extreme by discretizing the continuum box on a one point 
lattice with $L=1$. The running of the coupling will be determined in this case 
from a step scaling procedure that uses the rank of the gauge group as a size parameter.  
Step scaling will proceed by scaling the gauge group from $SU(N)$ to $SU(s^2 N)$.
The continuum step scaling function is thus obtained from the extrapolation
\be
\sigma(u,s) = \lim_{N\rightarrow \infty} \Sigma(u,s,\sqrt{N})\, ,
\ee
at fixed $u$. Here we have used that $\tilde l=a\sqrt{N}$ gives the effective lattice size, and thus for fixed $\tilde l$
the continuum limit is approached by sending $N$ to infinity.
The TGF coupling is automatically $\mathcal{O}(a)$ improved \cite{Ramos:2014kka,sintramos2} thus an extrapolation in $a^2 \sim 1/N$
will be required.

\section{Perturbative analysis of the twisted gradient flow on the lattice}
\label{s.disc}

Before presenting the outcome of the step scaling analysis we need to discuss the lattice 
definition of the TGF coupling.
We will just summarize the main results; all the technical details are included in Appendix \ref{ap.a}.
Let us recall that we are discussing the case of $SU(N)$ gauge theories discretized on an $L^4$ lattice
with twisted boundary conditions. The discussion will be done for arbitrary $L$, the TEK case follows easily by setting $L=1$. 
One has to start by considering a discretization of the flow equation and the lattice action used in the
Monte Carlo simulation.  We will focus on the case in which the Wilson plaquette action is used for both. 
For our choice of twist tensor Eq. (\ref{eq.twist}), it reads:
\be
S = b N \sum_n \sum_{\mu \nu } [N - Z_{\mu \nu }(n) \Tr (U_\mu(n) U_\nu(n+\hat \mu ) U_\mu^\dagger(n+\hat \nu ) U_\nu^\dagger(n)) ]\, .
\ee
with $Z_{\mu \nu }(n)=1$ for all plaquettes except for one corner plaquette in each plane, for which:
\be
Z_{\mu \nu }(n) = \exp \Big \{ i \, { 2 \pi k \over \sqrt{N}} \Big \}\, , \ \ {\rm for } \, \mu < \nu \, ; \ \  Z_{\nu \mu }(n) = Z_{\mu \nu }^*(n) .
\ee

The next step is to consider lattice approximants to the observable $E(t)$.
There are two standard choices in the literature: the plaquette definition
\be
\label{Eplaq}
\frac{t^2 E_P(t)}{N} =  \frac{t^2}{N} \Tr (1- \langle Z_{\mu \nu }(n) P_{\mu \nu}(n,t) \rangle)\, ,
\ee
where 
\be
P_{\mu \nu}(n,t) \equiv U_\mu(n) U_\nu(n+\hat \mu)  U_\mu^\dagger(n+\hat \nu) U_\nu^\dagger(n)\, , 
\ee
and the symmetric one
\be
\label{Esim}
\frac{t^2 E_S(t)}{N} =  \frac{t^2 }{2N} \langle \Tr \Big(\widehat G_{\mu \nu} (n,t) \widehat G_{\mu \nu} (n,t)\Big ) \rangle \, ,
\ee
where
\bea
\widehat G_{\mu \nu}(n,t) &\equiv& - \frac{i}{8} \{ Z_{\mu \nu }(n)  P_{\mu \nu}(n,t) +  Z_{\mu \nu }(n-\hat \nu) P_{-\nu \mu}(n,t) \\
&+&   Z_{\mu \nu }(n-\hat \mu) P_{\nu -\mu}(n,t)+   Z_{\mu \nu }(n-\hat \mu - \hat \nu) P_{-\mu -\nu}(n,t) -c.c.\} \, , \nonumber
\eea
and $U_{-\mu} (x) \equiv U_\mu^\dagger (x-\mu)$.

To have an idea of the artifacts induced by the discretization we can compare the lattice and the continuum
definitions of $t^2 E(t)/N$ at tree-level in perturbation theory. 
The lattice expressions are derived in Appendix  \ref{ap.a}. 
We obtain:
\bea
\label{eq.plaq}
\frac{t^2 E_P}{N}\Big |_{\rm tree}  &=&   \frac{3 \lambda_0 t^2}{2 N^2 L^4 } \sum'_q e^ {-2t  \widehat q^2}\, ,
\\
\label{eq.sym}
\frac{t^2 E_S}{N}\Big |_{\rm tree}  &=&   \frac{\lambda_0 t^2}{2 N^2 L^4} \sum_{\mu\ne \nu}\sum'_q e^ {-2t \widehat q^2} 
\, \,  \sin^2 (q_\nu) \cos^2 (q_\mu/2)  \, \frac{1}{\widehat q^2}\, .
\eea
The lattice momentum $ \widehat q_\mu = 2 \sin(q_\mu/2)$, where $q_\mu$ is given by:
\be
q_\mu = \frac{2 \pi m_\mu}{L \sqrt{N}}\, ,
\ee
with $m_\mu  = 0, \cdots,  L \sqrt{N}-1$. 
 The comparison between the different tree-level expressions for $t^2 E(t)/N$ is displayed on Fig. \ref{f.flow}. 
The dependence on the flow time $t$ of the
plaquette (blue) and  symmetric (red) definitions compared to the continuum expression (green) is displayed in 
\crefformat{figure}{Fig.~#2#1{(a)}#3} \cref{f.flow}.
Reduced lattice artifacts are observed
for the symmetric definition. Note however that this effect is strongly dependent on the lattice action used
in the flow equation \cite{Ramos:2014kka,sintramos2}. For example,  substituting in Eqs. (\ref{eq.plaq}) and (\ref{eq.sym}) the lattice kernel, 
$\exp(-2t  \widehat q^2)$, by the continuum one, $\exp(-2tq^2)$,
we obtain the results displayed in \crefformat{figure}{Fig.~#2#1{(b)}#3} \cref{f.flow}. In this case the plaquette definition
approximates the continuum result much better than the symmetric one. For comparison we also display in \crefformat{figure}{Fig.~#2#1{(c)}#3} \cref{f.flow}
the results that are obtained if the Symanzik improved Square action \cite{Snippe:1996bk,Snippe:1997ru} is used for the flow
(details are given in Appendix  \ref{ap.a}).

\FIGURE[ht]{
\centerline{
\includegraphics[angle=0,width=5.0cm]{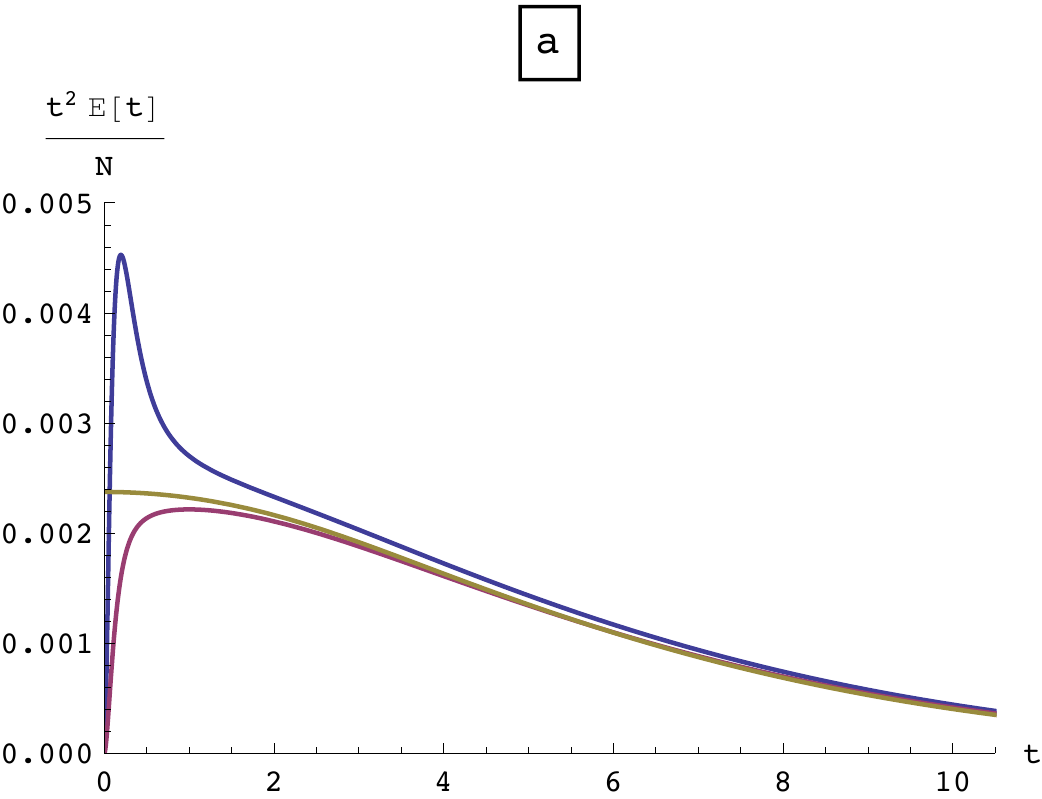}
\includegraphics[angle=0,width=5.0cm]{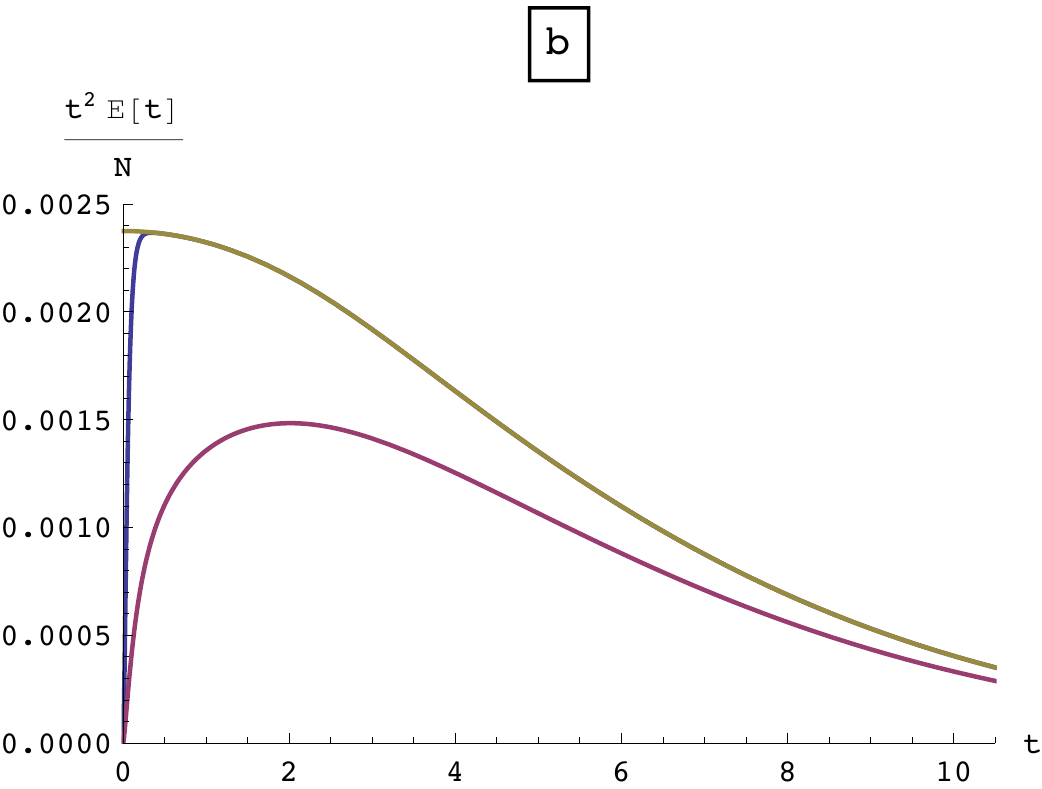}
\includegraphics[angle=0,width=5.0cm]{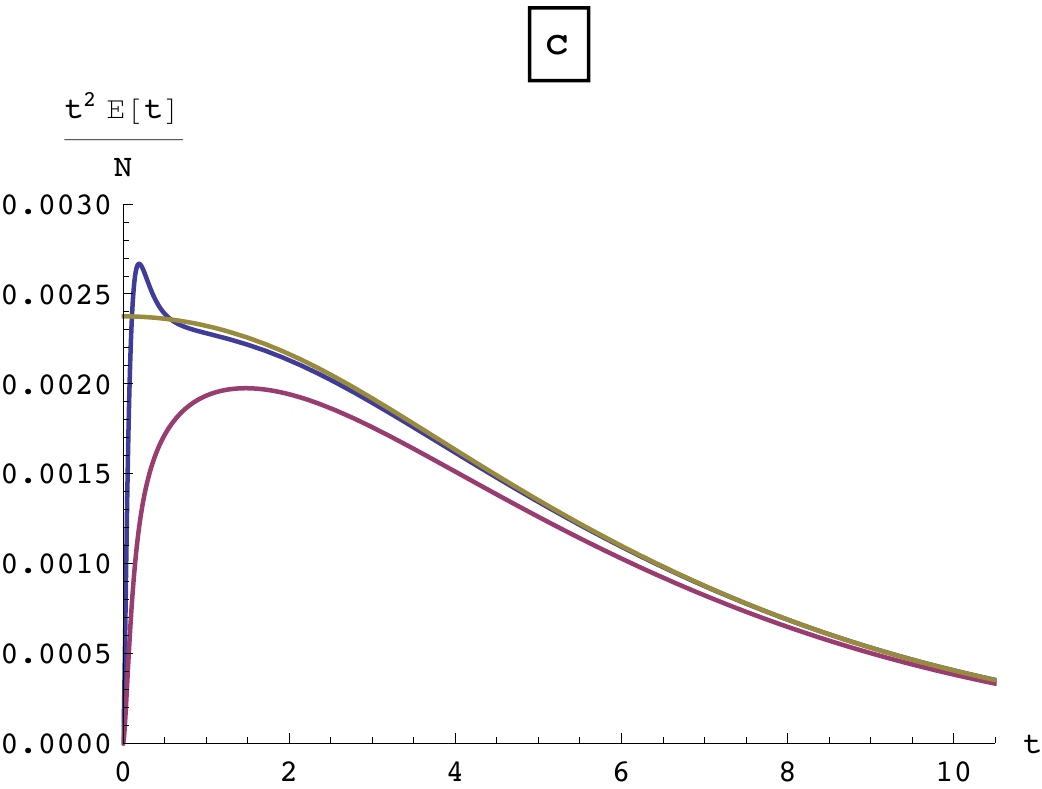}}
\caption{
\label{f.flow}
We display the tree-level perturbative dependence of $t^2 E/N$ on the flow time $t$ for $L=1$ and $\sqrt{N}=13$.
Results for the plaquette (blue) and  symmetric (red) definitions are compared to the continuum
expression (green). The different plots correspond to different kernels inserted in the flow equation:
{\bf (a)} the Wilson lattice kernel $ \exp (-2t \widehat q^2)$;  {\bf (b)} the continuum kernel $ \exp (-2t q^2)$; {\bf (c)} the
kernel of the Symanzik improved Square action $ \exp (-2t \tilde q_\mu^2)$, with $\widehat q$ and $\tilde q_\mu$
given by Eqs. (\ref{eq.qhat}) and (\ref{eq.qtilde}) respectively.}
}

These artifacts affect the determination of the TGF running coupling. 
A significant improvement is obtained if one adjusts the normalization constant $\mathcal{N}(c)$ entering the definition of the coupling
to preserve the equality between renormalized and bare coupling at leading order on the lattice
\cite{Fritzsch:2013je,Ramos:2014kla,Fodor:2014cpa,Perez:2014xga}. 
For our purposes we will only need $\mathcal{N}(c)$ on the TEK single-site lattices when the  Wilson 
plaquette action is used  
for the simulation and the flow. From Eqs. (\ref{eq.plaq}) and (\ref{eq.sym}) we derive
\be
\label{Nplaq}
\mathcal{N}_{P}(c) =  \frac{3  c^4}{128} \sum'_q  e^ {- c^2 N  \widehat q^2 /4}
\ee
and
\be
\label{Nsym}
\mathcal{N}_{S}(c) =  \frac{ c^4}{128}  \sum_{\mu\ne \nu}\sum'_q e^ {-c^2 N   \widehat q^2/4}
\, \,  \sin^2 (q_\nu) \cos^2 (q_\mu/2)  \, \frac{1}{\widehat q^2}\, ,
\ee
depending on whether the plaquette or symmetric definition of the coupling is employed.
If $\mathcal{N}(c)$ is chosen appropriately for each observable a significant reduction in lattice artifacts is achieved. Examples
will be presented in sec. \ref{s.results}. Most of the results that will be discussed in the next 
section correspond to these improved coupling definitions.

\TABLE{
\begin{tabular}{c|c|c|c|c|c}
& $\sqrt{N}=8$ & $\sqrt{N}=10$ & $\sqrt{N}=12$ & $\sqrt{N}=15$ & $\sqrt{N}=18$\\
\hline
$(k, |\bar k|)$ & (3,3) & (3,3) & (5,5) & (4,4) & (5,7)\\
$\tilde\theta/2\pi = |\bar k| / \sqrt{N}$ & 0.375 & 0.300 & 0.417 & 0.267 & 0.389 \\
\end{tabular}
\caption{\label{tab:k} Run parameters for each $N$.
The flux, denoted by $k$, is an integer coprime with $\sqrt{N}$.  $\tilde\theta$ equals $2 \pi |\bar k| / \sqrt{N}$, with
$\bar k$ an integer satisfying $k \bar k = 1$ (mod $\sqrt{N}$).}
}

\section{Results}
\label{s.results}

In this section we will compute the non-perturbative running coupling in the TEK model following the steps described in the previous sections. 
The procedure involves a numerical determination of the lattice step scaling function $\Sigma(u,s,\sqrt{N})$. 
Ideally one would start by measuring the TGF coupling on a set of $SU(N)$ TEK lattices,  
tuning the bare coupling $b$ to obtain the same value of the renormalized coupling $u$ for several values of $N$:
\be
u=\lambda_{\rm TGF}(\sqrt{N},b_N(u)) \, .
\ee
A fixed value of $u$ determines the line of constant physics.
For a given $N$ and scale factor $s$, the lattice step scaling function $\Sigma(u,s,\sqrt{N})$ is given by:
\be
\Sigma(u,s,\sqrt{N}) = \left.\lambda_{\rm TGF}(s\sqrt{N},b_N(u))\right|_{\lambda_{\rm TGF}(\sqrt{N},b_N(u))=u}\, ,
\ee
with the new renormalized coupling computed on a $SU(s^2 N)$ TEK lattice at the same value of the bare coupling $b_N(u)$.
The continuum step scaling function is obtained from the extrapolation $N\rightarrow \infty$ at fixed $u$.
This step is iterated several times, starting each run from a new value of $u_{n+1}=\sigma(s,u_n)$. 

The need to tune $b$ for each step makes this approach computationally expensive, and since the continuum extrapolation has 
to be taken for each step before repeating the tuning of $b$ for the next step, each step requires a new set of simulations.

A more economic alternative consists of measuring the TGF 
renormalized coupling for a wide range of values of $b$ at each value of $N$, and making use of an interpolating function 
to extract $\Sigma$ at any desired fixed value of $u$ from this data. This is the approach that we have followed in this paper.

\subsection{Simulation details}

The lattice action employed in the Monte Carlo simulation is the TEK model action given by Eq. (\ref{eq.tek}).
We generate 2000 configurations for a range of values of $b$, at $\sqrt{N}=8,10,12,15,18$.
This allows us to determine $\Sigma(u,s,\sqrt{N})$ for $s=3/2$ at three values of the lattice spacing 
($\sqrt{N}\equiv \lt/a = 8 \rightarrow 12$, $10 \rightarrow 15$, $12 \rightarrow 18$). 
Each configuration is separated by 250--1600 sweeps, where each sweep consists of one 
heat--bath and 5 over--relaxation updates, such that autocorrelations between configurations in the measured coupling are negligible.
The specific run parameters for each value of $N$ are listed in table \ref{tab:k}.

\FIGURE[ht]{\centerline{
\includegraphics[angle=270,width=7.0cm]{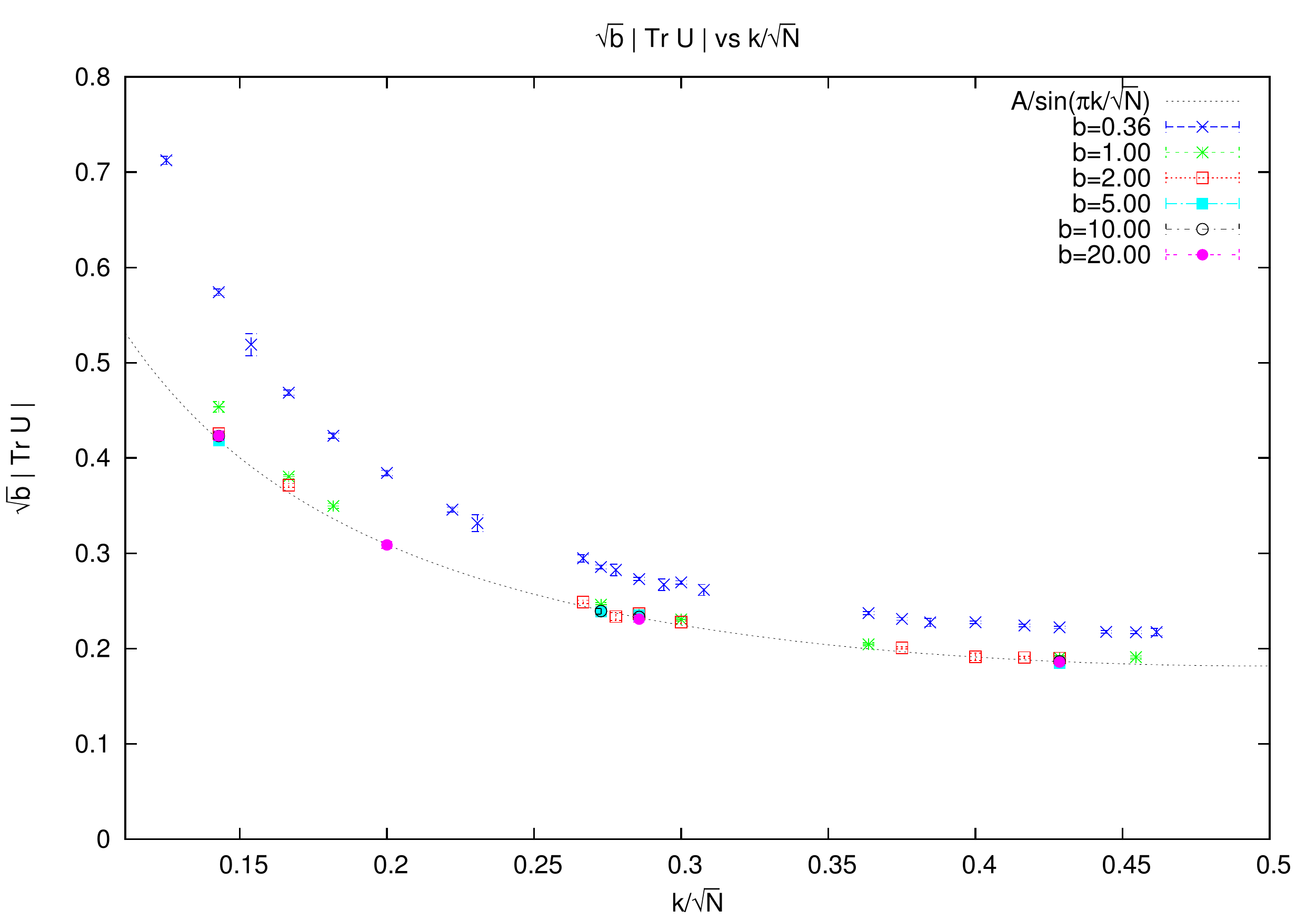}
\includegraphics[angle=270,width=7.0cm]{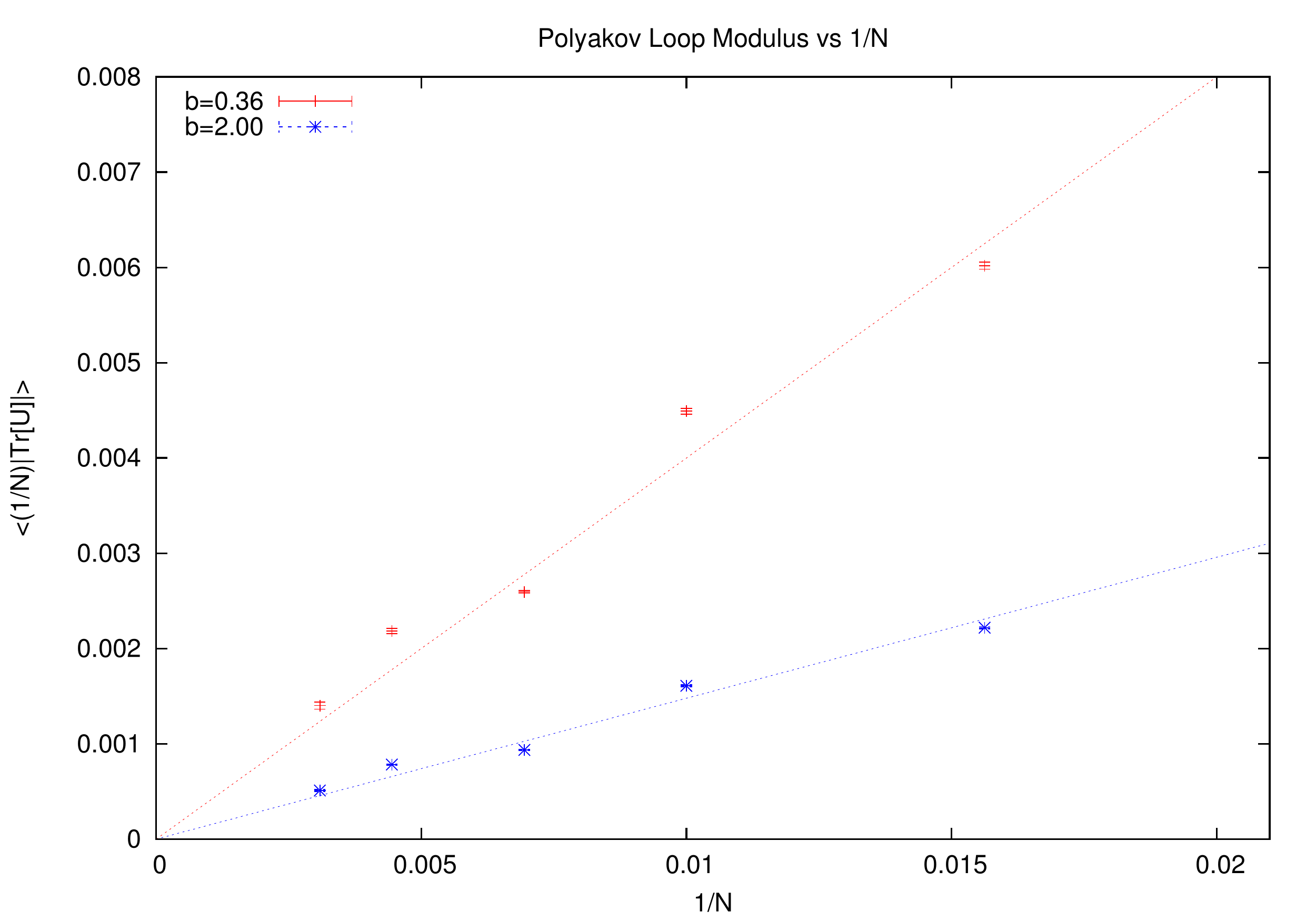}
}
\caption{\label{fig:poly} {\bf (a)} $\sqrt{b} \left| \Tr U_{\mu} \right|$ vs $k/\sqrt{N}$
for many values of $N$ and $b$, along with the perturbative prediction: $\sqrt{b}\left |\Tr(U_\mu) \right | = \left |A/\sin(\pi k/\sqrt{N})\right | $;
{\bf (b)} $\tfrac{1}{N}\left| \Tr U_{\mu} \right|$ vs $1/N$ at the strongest
and weakest couplings used in this work. The scatter of the points reflects the dependence on 
$k/\sqrt{N}$. }}

Before presenting the results for the coupling it is convenient to make a few comments regarding 
the validity of reduction for our set of lattices. There are certain restrictions on the 
allowed values of the flux $k$ and the integer $\bar k$.  
Let us briefly describe what they are.

Following  \cite{GonzalezArroyo:2010ss},  center symmetry is preserved on the TEK lattices if 
$k/\sqrt{N} > 1/9$. As an example of the behaviour of Polyakov loops, which act as order parameters for 
center symmetry breaking, we have analyzed the quantity $\left|\Tr U_{\mu} \right|/N$. 
\crefformat{figure}{Fig.~#2#1{(a)}#3} \cref{fig:poly} shows the               
quantity $\sqrt{b}\left|\Tr U_{\mu} \right|$ as a function of $k/\sqrt{N}$ for many values of $N$ and $b$.
For a given value of $b$, the points lie on a single curve. Furthermore, for 
$b\gtrsim 1$ there is no dependence on $b$.
These results are in qualitative agreement 
with the perturbative prediction.  To see this, take into account that the quantity measures
$\Tr(A_\mu \Gamma_\mu)\propto \sqrt{N}\hat{A}_\mu(p)$ for a particular value of
the momentum $p$. The actual value of $p$ depends on the flux $k$. It is given
by $p=(0,p_c,p_c,p_c)$ with $p_c=2 \pi k/\sqrt{N}$. In perturbation theory $\hat{A}_\mu(p)$ has a Gaussian distribution with a
width $\sigma(p) \sim 1/(\sqrt{Nb}|p|)$. For dimensional reasons the expectation value of  $|\hat{A}_\mu(p)|$
is also proportional to $\sigma(p)$. 
Replacing the continuum momentum by lattice momentum, our considerations lead 
to $\sqrt{b} |\Tr(U_\mu)|= |A/\sin(\pi k/\sqrt{N})|$ which describes the data very well for $b\gtrsim 1$.
This quantity divided by $\sqrt{b} N$ is shown as a function of $1/N$ for the weakest and strongest values of the coupling 
used in this work in \crefformat{figure}{Fig.~#2#1{(b)}#3} \cref{fig:poly}. By definition it is always positive but 
the figure shows that it goes to zero in the large $N$ limit.

\TABLE{
\begin{tabular}{c|c|c|c|c}
$\tilde \theta / 2\pi = |\bar{k}|/\sqrt{N}$  & 2/11 & 3/11 & 4/11 & 5/11 \\
\hline
$b= 1.00$ & 1.005(2) & 1.003(2) & 1.002(2) & 1 \\
$b= 0.36$ & 1.121(6) & 1.029(5) & 1.005(5) & 1 \\\end{tabular}
\caption{\label{t.kbar}The dependence of the coupling $\lambda_{\TGF}$ (normalized to the value at $\bar k =5$) on the quantity $\tilde \theta / 2\pi =
|\bar{k}|/\sqrt{N}$, for $\sqrt{N}=11$.}}

At finite values of $N$, the results obtained for the renormalized coupling will depend also on 
the value of the quantity $\tilde \theta$. A smooth continuum limit is best obtained by taking large 
$N$ while keeping the value of this quantity fixed. Strictly speaking this is impossible since 
$\bar k$ and $\sqrt{N}$ are coprime integers, introducing a source of systematic errors in our data. 
Nevertheless, it is to be expected that if $\tilde \theta$ is taken approximately constant this 
effect would be small. To test this question we present in table~\ref{t.kbar} the dependence on 
$\tilde \theta$ of the TGF coupling for $N=121$ and two values of $b$. One sees that at weak 
coupling this dependence is negligible.
At strong coupling it can become a sizable effect. Notice, however, that if $\tilde \theta > \pi/2$
the effect is at most 3 $\%$. This explains the values of $\tilde \theta$  used in our analysis 
and given in table \ref{tab:k}.

\subsection{Step Scaling Function}

 To determine the TGF running coupling, we integrate the gradient flow using the 3rd order Runge--Kutta 
scheme proposed in Ref. \cite{Luscher:2010iy} with an integration step-size $\Delta t$ in the range $0.01-0.03$, 
such that integration errors are much smaller than the statistical uncertainties.
We have computed the tree-level improved couplings determined from either the plaquette or the symmetric 
definition using the lattice determined constants $\mathcal{N}_{P}(c)$ or $\mathcal{N}_{S}(c)$ given in 
Eqs. (\ref{Nplaq}) and (\ref{Nsym}) respectively. 
The parameter $c$ is in principle free, and different values correspond to different renormalization 
schemes. In general, a smaller value of $c$ will result in smaller statistical uncertainties, but at the cost of 
larger lattice artifacts, and vice versa \cite{Fritzsch:2013je}. Here we take $c=0.30$ as a good compromise 
between these two effects. The measured couplings using the plaquette and symmetric definitions are listed in 
Appendix \ref{ap.b} in tables~\ref{tab:EP} and \ref{tab:ES} respectively. They have statistical errors $\mathcal{O}(0.3 -0.5\%)$.
For $\sqrt{N} \geq 10$ there is no clear dependence on the choice of discretisation within
the statistical errors.

\subsection{Continuum Extrapolation}

\FIGURE[ht]{
\centerline{\includegraphics[angle=0,width=8.0cm]{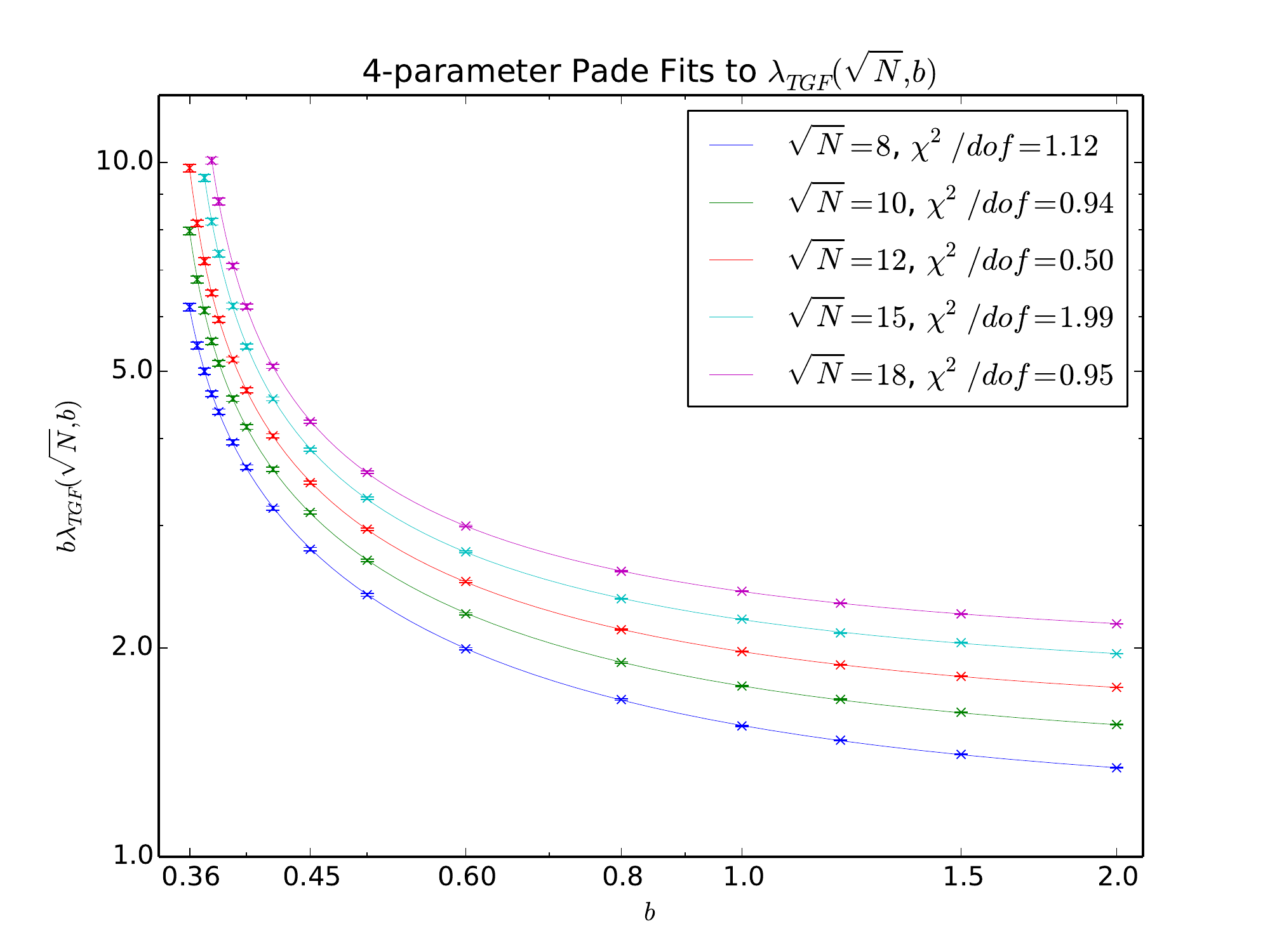}\hspace*{-0.2cm}\includegraphics[angle=0,width=8.0cm]{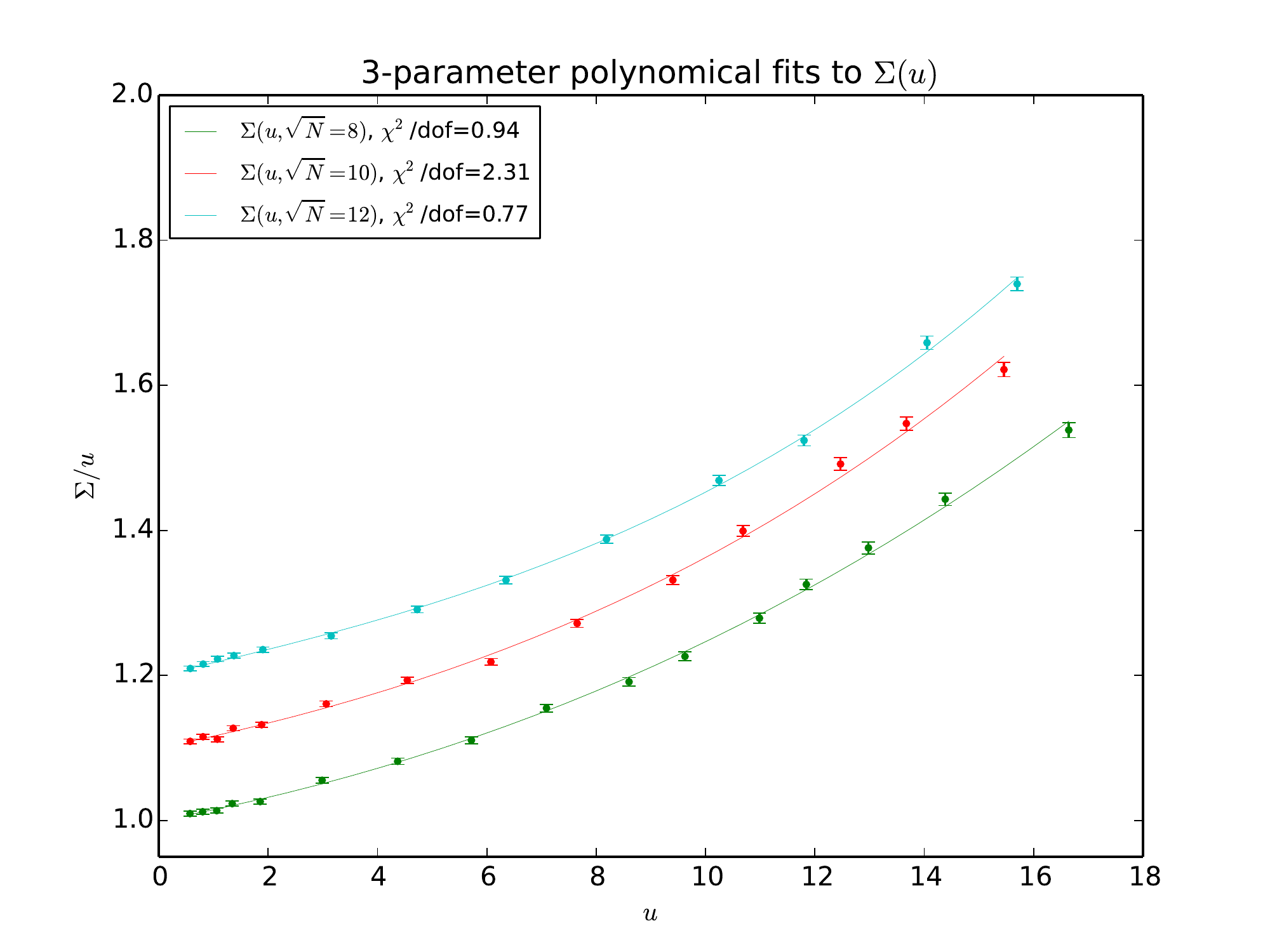}}\caption{\label{fig:pade}
Left: Dependence of $b \lambda_{\TGF}(\sqrt{N},b)$ on the bare coupling $b$ for various values of $N$. The lines represent 4--parameter Pad\'e interpolating fits to the data.
Right: Dependence of $\Sigma(u,s=3/2,\sqrt{N})/u$ on $u$. The lines represent 3--parameter polynomial interpolating fits to the data.
The data corresponding to different $N$ values have been displaced vertically for clarity (by 0.2 and 0.1 for the left and right--hand plots
respectively).}}

The results of the lattice step scaling function have to be extrapolated to the continuum limit at a fixed value of the
renormalized coupling $u$. 
%
%
In order to do that we have to interpolate the data. We use two different interpolating strategies. The first, 
following Ref. \cite{Fritzsch:2013je}, is to fit 
the $b$ dependence of the coupling to a 4--parameter Pad\'e ansatz of the form: 
\begin{equation}
\label{eq.pade}
\lambda_{\TGF}(\sqrt {N},b) = \frac{1}{b}\, \frac{a_0 + a_1 b + b^2}{a_2 + a_3 b + b^2} \,.
\end{equation}
Examples of such fits for the symmetric definition of the coupling are displayed in the left--hand plot of
Fig. \ref{fig:pade}. For plotting purposes the quantity plotted is $b \lambda_{\TGF}(\sqrt{N},b)$ and data corresponding to different values of 
$N$ have been displaced vertically by 0.2.
We obtain good fits with typical $\chi^2$ per degree of freedom of order 1. The Pad\'e fits allow us to extract the lattice step scaling function 
for arbitrary values of $u$.

\FIGURE[ht]{
\centerline{
\includegraphics[angle=270,width=7.0cm]{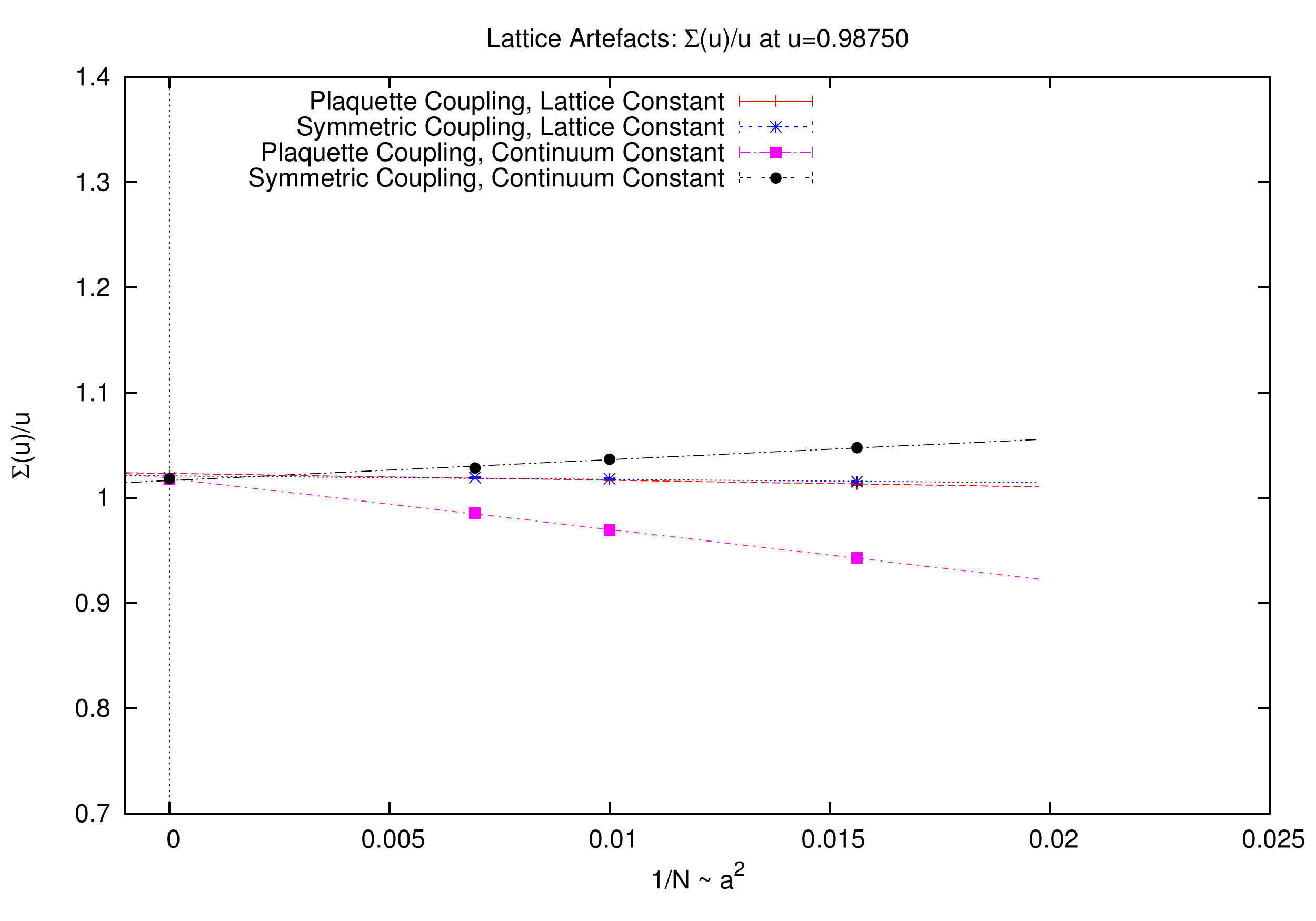}\includegraphics[angle=270,width=7.0cm]{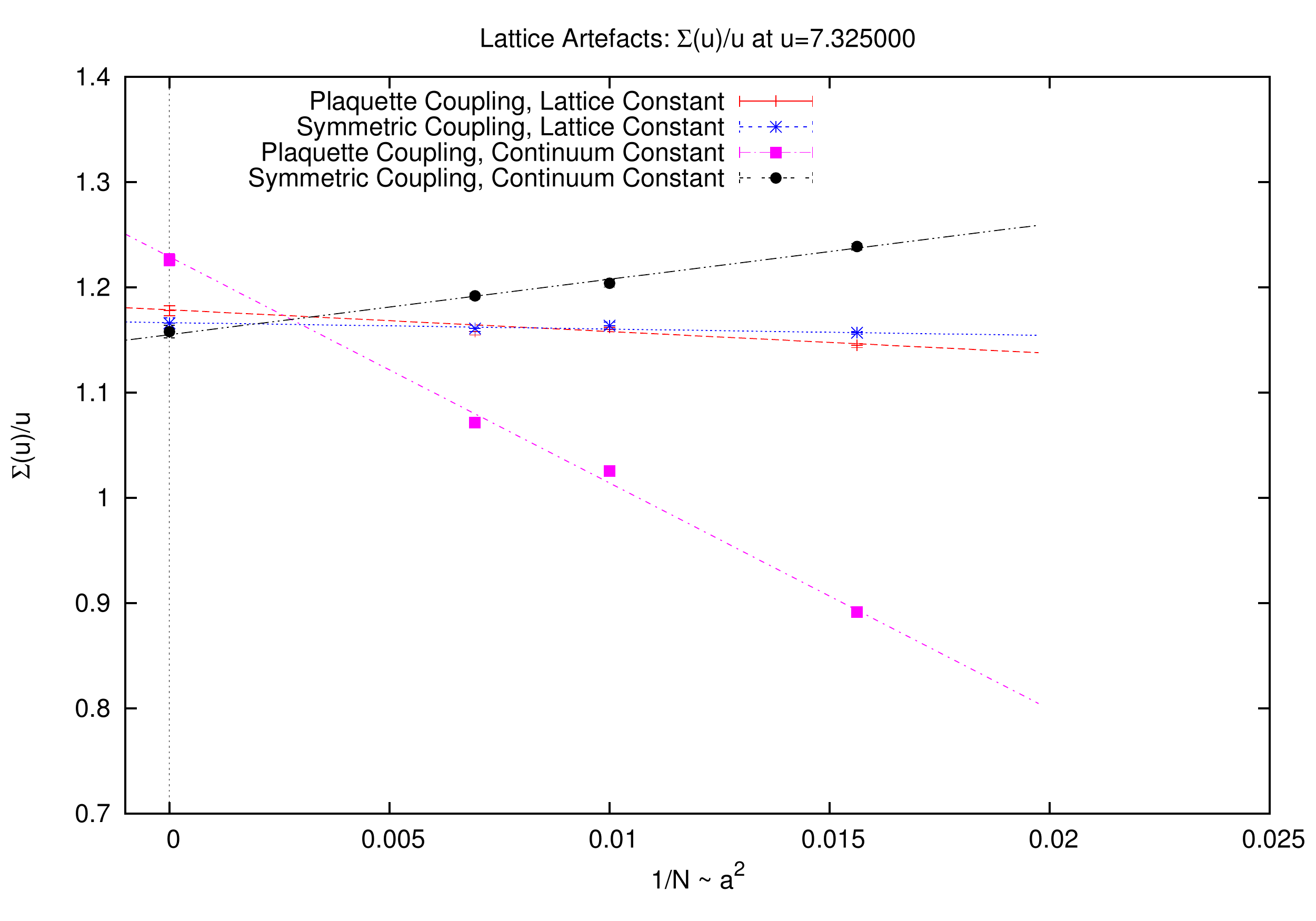}
}
\caption{
\label{f.arte}We display the continuum extrapolation of the step scaling function $\Sigma(u,s=3/2,\sqrt{N})/u$ for  $u=0.9875$ (Left) and
 $u=7.3250$ (Right).
The plots  illustrate the size of lattice artifacts depending on the choice of normalization constant $\mathcal{N}(c)$.}
}

\FIGURE[ht]{\centerline{
\includegraphics[angle=0,width=11.0cm]{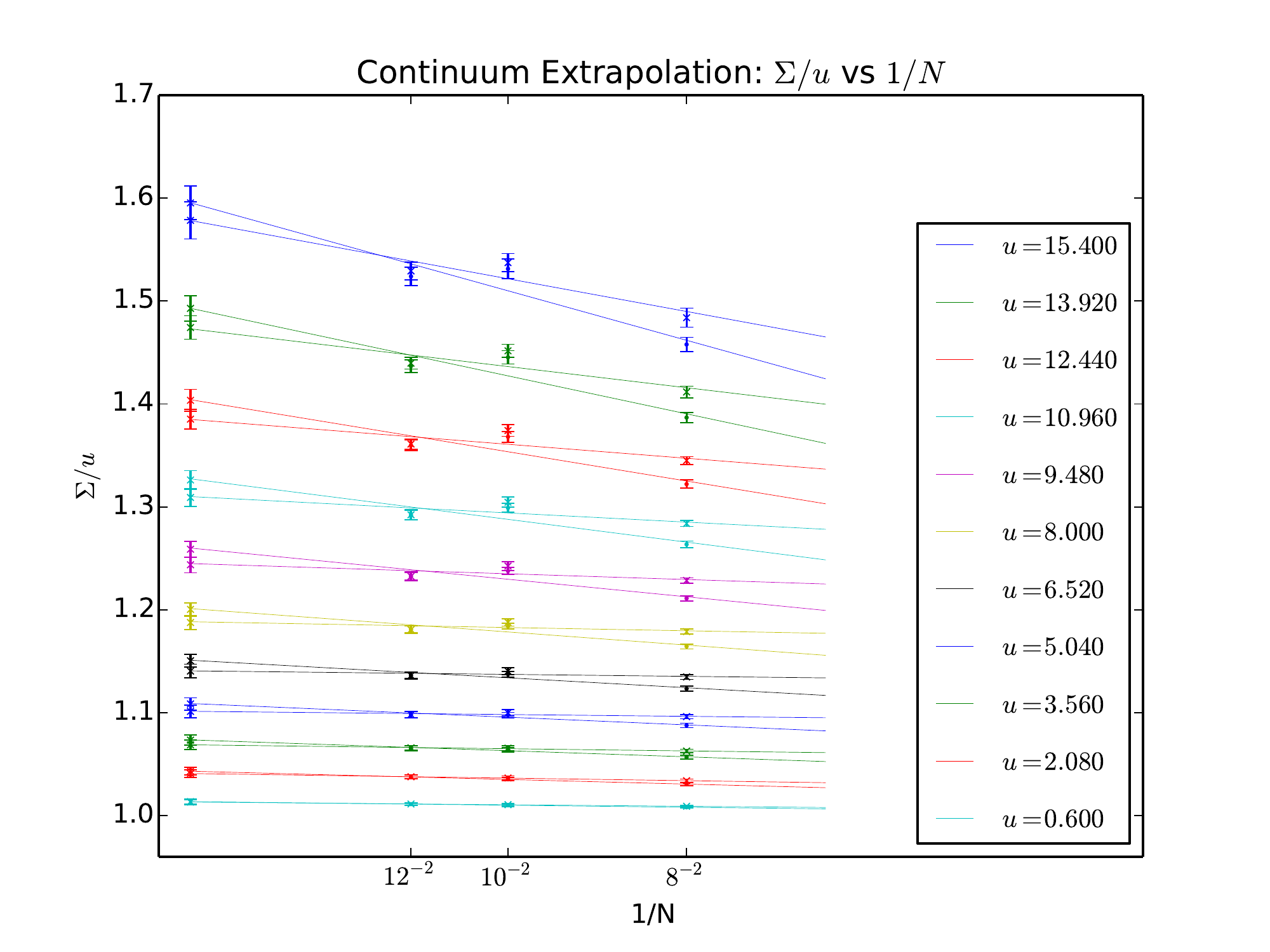}}
 \caption{\label{fig:continuum} We show the continuum extrapolation of $\Sigma(u,s=3/2,\sqrt{N})/u$ for several 
representative values of $u$. The results correspond to the TGF scheme with $c=0.3$, using both
the plaquette (circles) and symmetric (crosses) definitions of the coupling.}}

\FIGURE[ht]{\centerline{
\includegraphics[angle=0,width=11.0cm]{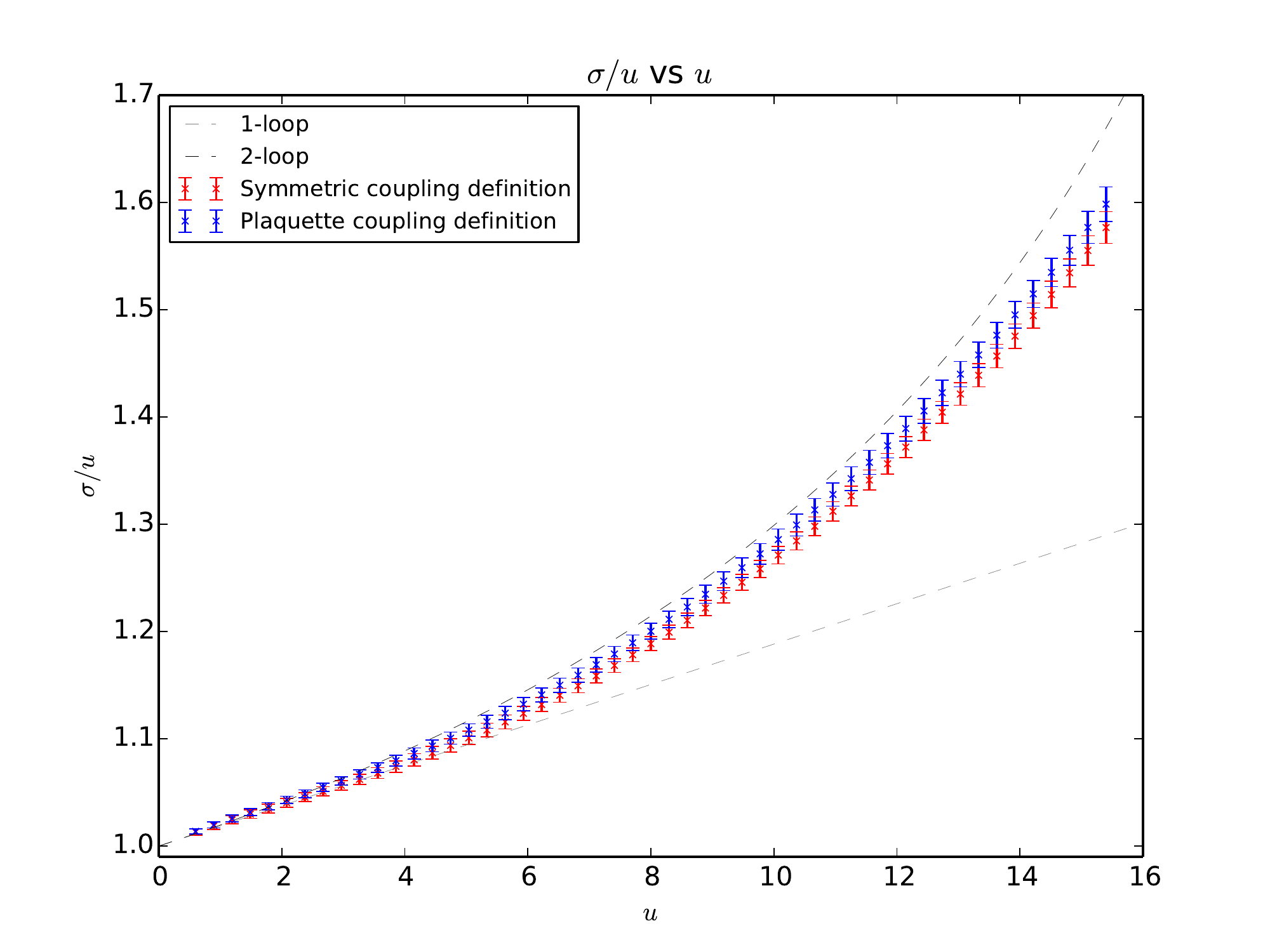}}
\caption{\label{fig:sigma_continuum} We display the continuum extrapolated step scaling function compared with the one-loop and two-loop
perturbative predictions. The results correspond to the TGF scheme with $c=0.3$.
}}

To check for systematic effects involved in the fitting procedure we also use a different strategy, where we first 
construct $\Sigma(u,s,\sqrt{N})/u$ directly from our coupling data, and
then interpolate this in $u$ using a 3--parameter polynomial of the form:
\begin{equation}
\label{eq.poly}
\Sigma(u,s,\sqrt{N})/u = 1 + a_0 u + a_1 u^2 + a_2 u^3 \,.
\end{equation}
Examples of these fits, along with the lattice data, are displayed in the right--hand plot of
Fig. \ref{fig:pade}, again for the symmetric definition of the coupling. The $\chi^2$ per degree of freedom for these fits are similar to those of the Pad\'e fits.

Let us start by illustrating the effect that the choice of $\mathcal{N}(c)$ has
on the size of lattice artifacts.  An example is presented in Fig. \ref{f.arte}.
We display the continuum extrapolation of the lattice step scaling function $\Sigma(u,s=3/2,\sqrt{N})/u$ for $u=0.9875$ and
 $u=7.3250$.
As already anticipated, the choice of the lattice definition of the renormalization constant results in a very significative decrease of lattice artifacts. 

Fig. \ref{fig:continuum} shows the continuum extrapolation of $\Sigma(u,s=3/2,\sqrt{N})$ for several representative 
values of $u$ ranging from $u=0.6$ to $u=15.4$. At each value of $u$ there are two different continuum extrapolations. Results obtained from the plaquette 
definition of the coupling are represented by circles, and those from the symmetric definition by crosses.
The difference between them at finite $N$ is a measure of lattice artifacts and they should give consistent continuum extrapolations.  
The analysis is repeated, using both types of interpolation, on a large number of bootstrap replicas of the data.
The central value and associated uncertainty are then determined from the mean and the variance of this set of bootstrap estimates.
Hence the error bars include both statistical errors, and the systematic error due to the choice of interpolation, 
although they do not include the systematic dependence introduced by not keeping $\tilde \theta$ exactly constant while taking the continuum limit. 
The fact that the $\sqrt{N}=10$ data is systematically higher than the rest might indeed be due to this effect. This source of error limits the accuracy 
of our continuum extrapolation which however does not seem to have a strong effect on the results.

The final, continuum extrapolated, result for $\sigma(u)/u$ is shown in 
Fig. \ref{fig:sigma_continuum} as a function of $u$, together with the 1--loop and 2--loop perturbative predictions. 

Our final result for the running coupling constant as a function of renormalization scale is presented in Fig. \ref{fig:running} and 
table \ref{tab:running} in Appendix \ref{ap.b}. 
We display $\lambda_{\TGF}(\tilde l)$ versus $\log_{3/2} (\tilde l/\tilde l_{\rm min})$ over a range of change in scale of $s^{30}$ with
$s=1.5$, starting at $\lambda_{\TGF}(\tilde l_{\rm max})=23.0$, 
and running down to $\lambda_{\TGF}(\tilde l_{\rm min})=1.65(10)$. A very good agreement with the 2--loop perturbative formula is observed at weak coupling.

\FIGURE[ht]{\centerline{
\includegraphics[angle=270,width=11.0cm]{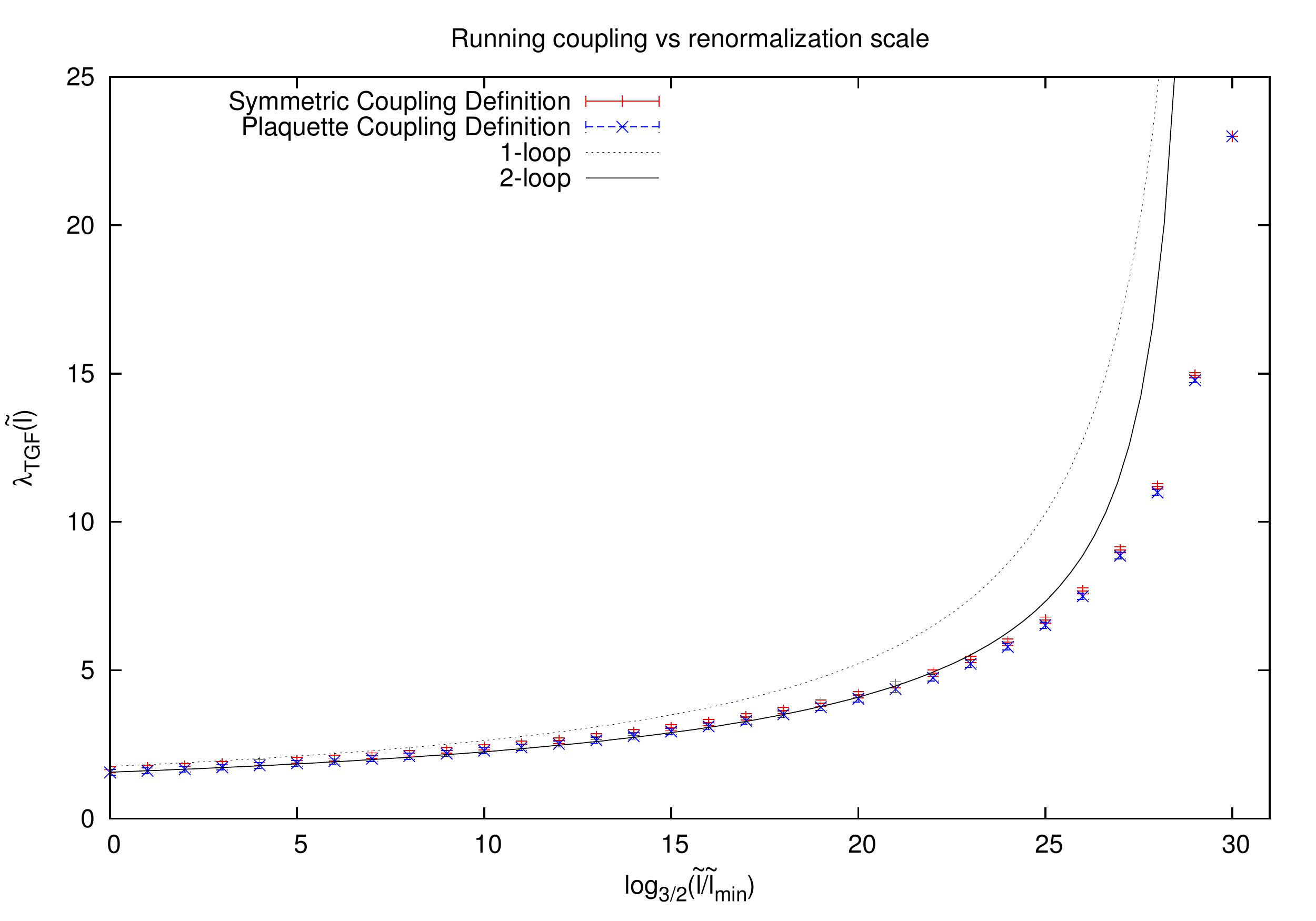}}
\caption{\label{fig:running} We display the continuum determination of the running coupling $\lambda_{\TGF}(\tilde l)$ versus $\log_{3/2} (\tilde l/\tilde l_{\rm min})$, along with the one-loop and two-loop          
perturbative predictions. The results correspond to the TGF scheme with $c=0.3$, starting at $\lambda_{\TGF}(\tilde l_{\rm max})=23.0$, and running 
down to $\lambda_{\TGF}(\tilde l_{\rm min})=1.65(10)$.}}

\section{Conclusions}

We have pushed the idea of volume independence to the extreme by determining the scale dependence of the $SU(N)$ renormalized gauge coupling 
from a scaling analysis of the single site TEK lattice, where the rank of the gauge group acts as a size parameter.
This allows us to determine the running of the coupling through a step scaling procedure that involves the scaling of the gauge
group $SU(N) \rightarrow SU(s^2 N)$. The continuum step scaling function is obtained in the $N\rightarrow \infty$ limit taken at fixed values
of the renormalized 't Hooft coupling. We have computed the running of the coupling across a wide range of scales, finding an excellent agreement with the
two-loop perturbative formula at weak coupling. Our results provide support to the conjecture that finite volume and finite $N$ effects are related
in the TEK model. 

To define the coupling we have proposed a modification of the Twisted Gradient Flow running coupling scheme introduced by A. Ramos in Ref. \cite{Ramos:2014kla}.
TGF is a finite volume renormalization scheme that uses the gradient flow \cite{Luscher:2010iy} combined with twisted boundary conditions to define 
the $SU(N)$ running coupling at a scale set by the size of the box. Our proposal, based on the idea of volume independence, has been to fix the 
renormalization scale in terms of an effective box size that combines finite volume and finite $N$ effects: $\tilde l = l \sqrt{N}$. The renormalized 
coupling at scale $\tilde l$ is thus given by:
\begin{equation}
\lambda_{\TGF}(\tilde l) = \left. \mathcal{N}^{-1}(c) \, \frac{t^2 \langle E (t) \rangle }{N} \right|_{t=c^2 {\tilde l}^2 /8} \, ,
\end{equation}
with $c$ an arbitrary constant parameter defining the renormalization scheme and $E(t)$ the energy density at a finite flow time $t$.
The proposal makes use of the idea of twisted volume reduction
conjecturing that finite volume effects on a 4-dimensional twisted box are controlled by the effective size parameter 
$\tilde l$ \cite{GonzalezArroyo:1982ub,GonzalezArroyo:1982hz,GonzalezArroyo:2010ss,Perez:2014sqa}.
This holds for a specific choice of twisted boundary conditions given by twist tensors $n_{\mu \nu}$ satisfying Eq.~(\ref{eq.twist}).  
For our choice of twisted boundary conditions, we have analyzed the tree-level perturbative behaviour of the energy density in the 
continuum and on the lattice. This allows to determine the 
normalization constant $\mathcal{N}(c)$ entering the definition of the running coupling. As already pointed out in Refs. 
\cite{Fritzsch:2013je,Ramos:2014kla,Fodor:2014cpa,Perez:2014xga}, the use of the lattice determined normalization constant results in a very significant 
reduction of cut-off effects in the running coupling. 

\appendix

\section{The normalization constant $\mathcal{N}(c)$ entering the definition of $\lambda_{TGF}$}
\label{ap.a}

In this section we will focus on the calculation of the normalization constant ${\cal N}(c)$ used
in the definition of the 't Hooft coupling within the Twisted Gradient Flow scheme given by Eq.~(\ref{lambda}).
As already mentioned, ${\cal N}(c)$ is determined by imposing that $\lambda_{TGF} (\tilde l)$ agrees at
tree-level with the bare coupling $\lambda_0$.
Although the results presented in this paper correspond to $d=4$ dimensions, in this Appendix we will keep
the discussion general by considering an arbitrary number of, even, space-time dimensions $d$.
In this case, the effective linear size of the box, in terms of which we fix the scale of the running coupling, is given by
$\tilde l = l N^{2/d}$.

To set the stage, we will start by deriving an  expression for ${\cal N}(c)$ in the continuum. 
Consider a $SU(N)$ gauge theory defined on a $d$ dimensional torus with periods $l$ and twisted boundary 
conditions given by the twist tensor $n_{\mu\nu}$. 
We will focus on the set of irreducible twist tensors~\cite{GonzalezArroyo:1997uj} given by:
\be
\label{eq.apptwist}
n_{\mu\nu} =\epsilon_{\mu \nu} \, k  N ^{d-2 \over d} \, ,
\ee
with
\be
\epsilon_{\mu\nu} =
\Theta(\nu-\mu)-\Theta(\mu-\nu)\, ,
\ee
with $\Theta$ the step function. For $d=4$ this reduces to the expression presented in Eq. (\ref{eq.twist}).
In this set up the perturbative formulas are derived by scaling the gauge potential as $g A_\mu$ and expanding all 
expressions in powers of the coupling $g$. The derivation will require us to consider the momentum expansion 
of the gauge potential compatible with the twisted boundary conditions:
\be
A_\mu(x) = {1\over l^{d/2}} \sum'_q  e^{i q x}  A_\mu (q) \HG(q)
\ee
with momenta quantized as:
\be
q_\mu = {2 \pi m_\mu \over \tilde l}\, , \, \, m_\mu \in Z\!\!\!Z\, ,
\ee
excluding those for which $q_\mu = 0$ (mod $N^{2/d}$) for all $\mu$ (indicated by the prime in the sum over momenta).
The matrices $\HG(q)$ are given by:
\be
\HG(q) = \frac{1}{\sqrt{2N}}\, e^{i \alpha(q)} \, \Gamma_0^{s_0(q)} \cdots \Gamma_{d-1}^{s_{d-1}(q)}
\, ,
\ee
with integers $s$ related to the momenta through:
\be
s_\mu (q) = \tilde \epsilon_{\nu\mu} \, \bar k \, q_\nu \, {\tilde l  \over 2 \pi}\, ({\rm mod} \, N^{2/d})\, .
\ee
The $\Gamma_\mu$ matrices are the so-called twist-eaters that satisfy:
\be
\Gamma_\mu \Gamma_\nu = \exp \{ 2 \pi i {n_{\nu \mu} \over N}\}  \Gamma_\nu \Gamma_\mu \, .
\ee
Here $\bar k$ is an integer defined through the relation:
\be
\label{bark}
k \bar k = 1 \, ({\rm mod} \, N^{2/d})\, ,
\ee
and $\tilde \epsilon_{\mu\nu}$ is an antisymmetric tensor satisfying:
\be
\sum_{\rho} \tilde \epsilon_{\mu\rho} \epsilon_{\rho \nu} = \delta_{\mu\nu} \, .
\ee

The flow equation can be solved order by order in perturbation theory. For that purpose it is
convenient to analyze the modified flow equation: 
\be
\label{eq.flowc}
\partial_t B_\mu (x,t,\xi) = D_\nu G_{\nu \mu} (x,t,\xi) + \xi D_\mu \partial_\nu B_\nu (x,t,\xi)   \, .  
\ee
Solutions of the original and modified flow equations are related by a flow-time dependent gauge transformation that
leaves $E(t)$ invariant.
At tree-level order, this is equivalent to solving the flow equation for the 
tree-level gauge fixed action.  In the Feynman gauge ($\xi=1$):
\be
S+ S_{GF} = - \int dx \Tr \{ A_\nu \partial_\mu^2  A_\nu \}\, .
\ee
A solution to the modified flow equation at this order is easily obtained in momentum space:
\be
B_\mu(x,t, \xi=1) = {1\over l^{d/2}} \sum'_q e^{- q^2 t } e^{i q x}  \hat B_\mu (q, t=0, \xi=1) \HG(q)\, .
\ee
Inserting this expansion in the expression for $t^2 E/N$ gives:
\be
\label{econt}
{t^2 E \over N}  =   {(d-1) \lambda_0 t^2  \over 2 {\tilde l}^d} \sum'_q  e^ {-2t q^2 }\, .
\ee
Through Eq. (\ref{lambda}), this leads to: 
\be
{\cal N}(c) =  {(d-1)  c^4 {\tilde l}^{(4-d)} \over 128} \sum'_q  e^ {- c^2 {\tilde l }^2  q^2 /4} = 
{(d-1)  c^4 {\tilde l}^{(4-d)} \over 128} \sum'_{m\in Z\!\!\!Z^d}  e^ {- c^2 \pi^2  m^2 }\, .
\ee
A compact expression for ${\cal N}(c)$ is obtained using the Jacobi Theta function:
\be
\theta_3(z,it) =  \sum_{m\in Z\!\!\!Z} e^ {- t \pi m^2 + 2 \pi i m z}
\ee
This gives the following expression
\be
{\cal N}(c) =  { (d-1)  c^4 {\tilde l}^{(4-d)} \over 128} \Big (\theta_3^d(0, i \pi c^2 )- \theta_3^d(0, i \pi c^2 N^{4/d} ) \Big)\, ,
\ee 
which can be easily evaluated.

Let us now proceed with the lattice calculation of ${\cal N}(c)$. 
For that purpose we discretize the $SU(N)$ gauge theory on a $L^d$ lattice endowed with twisted boundary conditions.
The torus periods are given by $l=L a$, with $a$ the lattice spacing. In what follows $a$ will be 
set to 1, $a$ dependent expressions can be easily retrieved by using dimensional arguments.
The expressions derived in this way will reduce to those of the TEK one-point lattice model by setting $L=1$.

As mentioned in section \ref{s.disc}, 
three ingredients have to be considered when deriving a lattice expression for $t^2 E(t)$:
\begin{itemize}
\item the discretized lattice action used in the Monte Carlo simulation,
\item the discretized lattice flow equation,
\item the discretization of the observable representing $E(t)$ on the lattice.
\end{itemize}
We will analyze the case in which the Wilson plaquette action is used both for the Monte Carlo simulation and for the flow. For twisted boundary 
conditions, it reads:
\be
S = b N \sum_n \sum_{\mu \nu } [N - Z_{\mu \nu }(n) \Tr (U_\mu(n) U_\nu(n+\hat \mu ) U_\mu^\dagger(n+\hat \nu ) U_\nu^\dagger(n)) ]\, .
\ee
with $Z_{\mu \nu }(n)=1$ for all plaquettes except for one corner plaquette in each plane, for which:
\be
Z_{\mu \nu }(n) = \exp \{ 2 \pi i {n_{\mu \nu} \over N} \}\, ,
\ee
$n_{\mu \nu}$ denoting the twist tensor given by Eq. (\ref{eq.apptwist}).

The gauge links are expanded in perturbation theory as:
\be
\label{eq.links}
U_\mu (n) =  V_\mu(n)  = e^{i g A_\mu(n + {\hat \mu \over 2})}\, ,
\ee
for all $n$ such that $n_\mu\ne L-1$, and
\be
\label{eq.links2}
U_\mu (n) =  V_\mu(n) \Gamma_\mu = e^{i g A_\mu(n + {\hat \mu \over 2})}  \Gamma_\mu\, ,
\ee
if $n_\mu=L-1$.

The modified lattice flow equation, equivalent to Eq. (\ref{eq.flowc}) in
the Feynman gauge $\xi=1$, is derived at tree-level from the discretized 
gauge fixed action:
\be
S + S_{GF} = - b N \sum_n \sum_{\mu \nu } \Tr \{ A_\nu(n) \nabla_\mu^- \nabla_\mu^+ A_\nu(n) \}\, ,
\ee
where the lattice forward and backward derivatives are given by:
\bea
\nabla_\mu^+ \phi = \phi(n+\hat \mu)  - \phi(n) \, ,\\
\nabla_\mu^- \phi = \phi(n) - \phi(n-\hat \mu)\, .
\eea
This gives at leading order in $g$:
\be
\partial_t B_\mu(n, t) = \nabla_\mu^- \nabla_\mu^+  B_\mu(n, t)\, ,
\ee
which is easily solved using the expansion in momenta of the gauge fields. The solution reads:
\be
\label{eq.bflow}
B_\mu(n, t) = {1 \over L^{d/2}} \sum'_q e^{- \widehat q^2 t } e^{i q (n+ {\hat \mu \over 2})} \hat B_\mu (q, t=0) \HG(q)\, ,
\ee
with lattice momenta 
\be
\label{eq.qhat}
 \widehat q_\mu = 2 \sin(q_\mu/2)\, , 
\ee
where $q_\mu$ is given by $ q_\mu = 2 \pi m_\mu /  \tilde L$ ,
with $m_\mu  = 0, \cdots, \tilde L-1$, and $\tilde L = L N^{2/d} $. 

In addition to the solution of the flow equation, one has to consider lattice aproximations to the observable
 $E(t)$.  
Using the Fourier expansion of the gauge potential, Eq. (\ref{eq.bflow}), and the lattice propagator 
for the Wilson action it is easy to derive the leading order expansions of the plaquette and symmetric 
definitions presented in Eqs. (\ref{Eplaq}) and (\ref{Esim}):
\be
{t^2 E_P \over N}  =   {(d-1) \lambda_0 t^2  \over 2 \tilde L^d } \sum'_q e^ {-2t  \widehat q^2}
\ee
\be
{t^2 E_S \over N}  =   {\lambda_0 t^2 \over 2  \tilde L^d} \sum_{\mu\ne \nu}\sum'_q e^ {-2t \widehat q^2}  
\, \,  \sin^2 (q_\nu) \cos^2 (q_\mu/2)  \, {1\over \widehat q^2}
\ee

 We are now ready to derive the lattice expressions for ${\cal N}$ obtained from the plaquette and 
symmetric observables. 
The condition to be imposed is that $\lambda_{TGF}$ in Eq. (\ref{lambda}) equals the bare coupling $\lambda_0$
at tree-level in perturbation theory. Taking into account that $\tilde l = \tilde L a$, this leads to:
\be
{\cal N}_P(c) =  {(d-1)  c^4 \tilde L^{4-d}\over 128} \sum'_q  e^ {- c^2 {\tilde L}^2  \widehat q^2 /4}
\ee
and
\be
{\cal N}_S(c) =  { c^4 \tilde L^{4-d}\over 128}  \sum_{\mu\ne \nu}\sum'_q e^ {-c^2 {\tilde L}^2   \widehat q^2/4}
\, \,  \sin^2 (q_\nu) \cos^2 (q_\mu/2)  \, {1\over \widehat q^2}
\ee

We have also analyzed the effect of lattice artefacts for other discretized versions of the flow equation. We have considered 
in particular the Symanzik improved Square action \cite{Snippe:1996bk,Snippe:1997ru} which combines $1\times 1$, $1\times 2$ and $2\times 2$ plaquettes:
\bea
S_{\rm sq} &=& b N \sum_{\mu \nu } \sum_n \{ c_0[N - Z_{\mu \nu }(n) \Tr P_{\mu  \nu} (n) ] \\
&+& 2 c_1 [N - Z_{\mu \nu }(n) Z_{\mu \nu }(n+\hat \mu) \Tr P_{(2\mu)  \nu}(n) ] \nonumber \\
&+& c_4 [N - Z_{\mu \nu }(n) Z_{\mu \nu }(n+\hat \mu) Z_{\mu \nu }(n+\hat \nu) 
Z_{\mu \nu }(n+\hat \mu+\hat \nu) \Tr P_{(2\mu)  (2\nu)}(n)]
\}\, ,
\eea
with tree-level improvement coefficients:
\be
c_0 = {16\over 9}, \, c_1 = - {1\over 9}, \, c_4 = {1\over 144}\,.
\ee
It has the advantage that, choosing appropriately the gauge fixing term, one can obtain a diagonal propagator.
This allows one to solve in a simple way the flow equation. The action at lowest order in $g$ reads:.
\be
S_{\rm sq}+S_{\rm GF} =
 {1 \over 18}  \sum'_q  A_\nu (-q)\Big (4-\cos^2(q_\nu/2)\Big ) \Big (4 \widehat q^2 -  \sum_\rho \sin^2(q_\rho)\Big )
    A_\nu (q)
\ee
The solution to the corresponding flow equation is given by:
\be
\label{eq.bflowsq}
B_\mu(n, t) = {1 \over L^{d/2}} \sum'_q e^{- 2 t \tilde q_\mu^2 } e^{i q (n+ {\hat \mu \over 2})} \hat B_\mu (q, t=0) \HG(q)\, ,
\ee
where
\be
\label{eq.qtilde}
\tilde q_\mu^2 = {1 \over 9} \Big (4 \widehat q^2 -  \sum_\rho \sin^2(q_\rho)\Big )
\Big (4-\cos^2(q_\mu/2)\Big ) \, .
\ee
The insertion of this expression into the plaquette and symmetric definitions of $E(t)$ leads to
\be
{t^2 E_P^{sq} \over N}  =   {\lambda_0 t^2  \over 2 {\tilde L}^d} \sum_\mu\sum'_q e^ {-2t  \tilde q_\mu^2} \Big(1- {\widehat q_\mu^2
\over \widehat q^2}\Big) \, ,
\ee
and
\be
{t^2 E_S^{sq} \over N}  =   {\lambda_0 t^2 \over 2 {\tilde L}^d} \sum_{\mu\ne \nu}\sum'_q e^ {-2t \tilde q_\mu^2}
\, \,  \sin^2 (q_\nu) \cos^2 (q_\mu/2)  \, {1\over \widehat q^2} \, .
\ee
The analysis of the lattice artefacts for the Square action is displayed in\crefformat{figure}{Figure~#2#1{(c)}#3} \cref{f.flow}.

\section{Numerical results for the running coupling constant}
\label{ap.b}

In Tables~\ref{tab:EP} and \ref{tab:ES}  we list the tree-level improved couplings determined respectively from
the plaquette and the symmetric definition.
The parameter $c$  has been set to $c=0.3$.  The results have statistical errors $\mathcal{O}(0.3 -0.5\%)$.

Our final results for the running coupling constant as a function of renormalization scale are listed in
Table \ref{tab:running}.

\TABLE{
\footnotesize
\begin{tabular}{c|c|c|c|c|c}
$b$ & $\sqrt{N}=8$ & $\sqrt{N}=10$ & $\sqrt{N}=12$ & $\sqrt{N}=15$ & $\sqrt{N}=18$ \\
\hline
0.360 & 16.801(72) & 20.981(97) & 25.53(12) & - & - \\
0.365 & 14.540(55) & 17.479(77) & 20.710(85) & - & - \\
0.370 & 13.121(48) & 15.442(64) & 17.830(78) & 23.44(11) & - \\
0.375 & 11.972(41) & 13.688(54) & 15.682(61) & 19.737(92) & 24.12(11) \\
0.380 & 11.101(36) & 12.487(48) & 14.044(54) & 17.313(79) & 20.459(96) \\
0.390 & 9.726(30) & 10.707(37) & 11.800(43) & 13.865(57) & 15.609(65) \\
0.400 & 8.684(26) & 9.424(31) & 10.247(36) & 11.573(43) & 12.990(53) \\
0.420 & 7.157(20) & 7.669(24) & 8.195(27) & 8.967(29) & 9.726(34) \\
0.450 & 5.763(15) & 6.089(18) & 6.356(18) & 6.798(20) & 7.186(24) \\
0.500 & 4.399(11) & 4.556(12) & 4.730(13) & 4.972(14) & 5.156(15) \\
0.600 & 3.0031(68) & 3.0688(74) & 3.1551(83) & 3.2510(84) & 3.3260(87) \\
0.800 & 1.8617(40) & 1.8840(45) & 1.9054(45) & 1.9425(46) & 1.9723(48) \\
1.000 & 1.3471(29) & 1.3630(30) & 1.3754(32) & 1.3991(34) & 1.4125(35) \\
1.200 & 1.0627(23) & 1.0716(24) & 1.0751(25) & 1.0845(25) & 1.0992(27) \\
1.500 & 0.8043(17) & 0.8104(18) & 0.8129(19) & 0.8227(19) & 0.8255(19) \\
2.000 & 0.5724(12) & 0.5755(12) & 0.5772(13) & 0.5806(13) & 0.5827(14)
\end{tabular}
\caption{\label{tab:EP} Measured coupling $\lambda_{\TGF}$ for each $b$ and $N$ (plaquette definition).}
}
\TABLE{
\footnotesize
\begin{tabular}{c|c|c|c|c|c}
$b$ & $\sqrt{N}=8$ & $\sqrt{N}=10$ & $\sqrt{N}=12$ & $\sqrt{N}=15$ & $\sqrt{N}=18$ \\
\hline
0.360 & 16.643(77) & 21.05(10) & 25.60(12) & - & - \\
0.365 & 14.383(61) & 17.492(82) & 20.755(88) & - & - \\
0.370 & 12.979(53) & 15.445(67) & 17.857(81) & 23.52(11) & - \\
0.375 & 11.843(45) & 13.672(58) & 15.698(63) & 19.788(94) & 24.17(11) \\
0.380 & 10.986(40) & 12.469(51) & 14.051(57) & 17.350(81) & 20.496(97) \\
0.390 & 9.624(33) & 10.685(40) & 11.801(44) & 13.882(59) & 15.626(66) \\
0.400 & 8.601(28) & 9.402(33) & 10.246(37) & 11.579(44) & 13.001(54) \\
0.420 & 7.091(22) & 7.652(25) & 8.190(28) & 8.966(29) & 9.727(35) \\
0.450 & 5.718(17) & 6.075(19) & 6.351(19) & 6.796(20) & 7.185(24) \\
0.500 & 4.370(12) & 4.546(13) & 4.726(14) & 4.971(14) & 5.156(15) \\
0.600 & 2.9878(76) & 3.0635(79) & 3.1536(87) & 3.2498(86) & 3.3256(88) \\
0.800 & 1.8556(46) & 1.8815(48) & 1.9041(47) & 1.9419(47) & 1.9720(49) \\
1.000 & 1.3434(33) & 1.3618(32) & 1.3747(33) & 1.3990(35) & 1.4123(36) \\
1.200 & 1.0603(26) & 1.0712(26) & 1.0747(26) & 1.0842(26) & 1.0990(27) \\
1.500 & 0.8030(19) & 0.8101(20) & 0.8127(20) & 0.8227(20) & 0.8255(20) \\
2.000 & 0.5716(13) & 0.5752(13) & 0.5771(14) & 0.5805(13) & 0.5826(14)
\end{tabular}
\caption{\label{tab:ES} Measured coupling $\lambda_{\TGF}$ for each $b$ and $N$ (symmetric definition).}
}
\TABLE{
\footnotesize
\begin{tabular}{c|c|c}
$\log_{3/2} (\tilde l_{\rm max}/\tilde l)$ & Plaquette coupling definition & Symmetric coupling definition \\
\hline
0 & 23.0 & 23.0 \\
1 & 14.776(75) & 14.950(79) \\
2 & 10.995(99) & 11.202(91) \\
3 & 8.865(112) & 9.062(100) \\
4 & 7.494(109) & 7.677(101) \\
5 & 6.522(107) & 6.693(101) \\
6 & 5.790(106) & 5.953(103) \\
7 & 5.216(107) & 5.372(104) \\
8 & 4.751(108) & 4.902(105) \\
9 & 4.365(109) & 4.511(105) \\
10 & 4.039(110) & 4.180(105) \\
11 & 3.759(110) & 3.895(105) \\
12 & 3.515(109) & 3.648(104) \\
13 & 3.300(109) & 3.429(104) \\
14 & 3.110(108) & 3.235(103) \\
15 & 2.939(107) & 3.061(102) \\
16 & 2.785(106) & 2.905(101) \\
17 & 2.645(105) & 2.762(100) \\
18 & 2.518(104) & 2.632(99) \\
19 & 2.401(103) & 2.514(98) \\
20 & 2.294(102) & 2.404(97) \\
21 & 2.195(101) & 2.303(97) \\
22 & 2.104(100) & 2.209(96) \\
23 & 2.019(99) & 2.122(96) \\
24 & 1.939(98) & 2.041(95) \\
25 & 1.866(98) & 1.966(95) \\
26 & 1.797(97) & 1.895(95) \\
27 & 1.732(96) & 1.828(95) \\
28 & 1.671(95) & 1.766(95) \\
29 & 1.614(95) & 1.707(95) \\
30 & 1.561(94) & 1.652(95) 
\end{tabular}
\caption{\label{tab:running} Running coupling $\lambda_{\TGF}(\tilde l)$ as a function of the scale $\tilde l$, for both the plaquette and symmetric  
definitions of the coupling. The parameter $c$ has been set to $c=0.3$.}
}

\FloatBarrier

\section*{Acknowledgments}
We are indebted to Alberto Ramos for many useful discussions on the TGF running coupling scheme.
We acknowledge financial support from the MCINN grants FPA2009-08785, FPA2009-09017, FPA2012-31686 and FPA2012-31880, the Comunidad Aut\'onoma de 
Madrid under the program HEPHACOS S2009/ESP-1473, the European Union under Grant Agreement PITN-GA-2009-238353 (ITN STRONGnet), and the 
Spanish MINECO's ``Centro de Excelencia Severo Ochoa'' Programme under grant SEV-2012-0249. 
M. O. is supported by the Japanese MEXT grant No 26400249. Calculations have been done on Hitachi SR16000 
supercomputer both at High Energy Accelerator Research Organization(KEK) and YITP in Kyoto University, and
the HPC-clusters at IFT.  Work at KEK is supported by the Large Scale Simulation Program No.14/15-03.

\bibliography{flow}

\providecommand{\href}[2]{#2}\begingroup\raggedright\begin{thebibliography}{10}

\bibitem{GonzalezArroyo:1982ub}
A.~Gonz\'alez-Arroyo and M.~Okawa, {\it {A Twisted Model for Large $N$ Lattice
  Gauge Theory}},  {\em Phys. Lett.} {\bf B120} (1983) 174.

\bibitem{GonzalezArroyo:1982hz}
A.~Gonz\'alez-Arroyo and M.~Okawa, {\it {The Twisted Eguchi-Kawai Model: A
  Reduced Model for Large N Lattice Gauge Theory}},  {\em Phys. Rev.} {\bf D27}
  (1983) 2397.

\bibitem{GonzalezArroyo:2010ss}
A.~Gonz\'alez-Arroyo and M.~Okawa, {\it {Large $N$ reduction with the Twisted
  Eguchi-Kawai model}},  {\em JHEP} {\bf 07} (2010) 043,
  [\href{http://arxiv.org/abs/1005.1981}{{\tt arXiv:1005.1981}}].

\bibitem{Gonzalez-Arroyo:2014dua}
A.~Gonzalez-Arroyo and M.~Okawa, {\it {Testing volume independence of SU(N)
  pure gauge theories at large N}},  {\em JHEP} (2014) {to appear},
  [\href{http://arxiv.org/abs/1410.6405}{{\tt arXiv:1410.6405}}].

\bibitem{Perez:2014sqa}
M.~Garc\'{\i}a~P\'erez, A.~Gonz\'alez-Arroyo, and M.~Okawa, {\it {Volume
  independence for Yang-Mills fields on the twisted torus}},  {\em
  International Journal of Modern Physics A} {\bf Vol. 29, No. 25} (2014)
  1445001, [\href{http://arxiv.org/abs/1406.5655}{{\tt arXiv:1406.5655}}].

\bibitem{Perez:2013dra}
M.~Garc\'{\i}a~P\'erez, A.~Gonz\'alez-Arroyo, and M.~Okawa, {\it {Spatial
  volume dependence for 2+1 dimensional SU(N) Yang-Mills theory}},  {\em JHEP}
  {\bf 1309} (2013) 003, [\href{http://arxiv.org/abs/1307.5254}{{\tt
  arXiv:1307.5254}}].

\bibitem{Perez:2012fz}
M.~G. Perez, A.~Gonzalez-Arroyo, and M.~Okawa, {\it {Volume dependence in 2+1
  Yang-Mills theory}},  {\em PoS} {\bf LATTICE2012} (2012) 219,
  [\href{http://arxiv.org/abs/1211.0807}{{\tt arXiv:1211.0807}}].

\bibitem{Luscher:1991wu}
M.~Luscher, P.~Weisz, and U.~Wolff, {\it {A Numerical method to compute the
  running coupling in asymptotically free theories}},  {\em Nucl. Phys.} {\bf
  B359} (1991) 221.

\bibitem{Narayanan:2006rf}
R.~Narayanan and H.~Neuberger, {\it {Infinite N phase transitions in continuum
  Wilson loop operators}},  {\em JHEP} {\bf 03} (2006) 064,
  [\href{http://arxiv.org/abs/hep-th/0601210}{{\tt hep-th/0601210}}].

\bibitem{Luscher:2010iy}
M.~Luscher, {\it {Properties and uses of the Wilson flow in lattice QCD}},
  {\em JHEP} {\bf 08} (2010) 071, [\href{http://arxiv.org/abs/1006.4518}{{\tt
  arXiv:1006.4518}}].

\bibitem{Ramos:2013gda}
A.~Ramos, {\it {The gradient flow in a twisted box}},  {\em PoS} {\bf
  Lattice2013} (2013) 053, [\href{http://arxiv.org/abs/1308.4558}{{\tt
  arXiv:1308.4558}}].

\bibitem{Ramos:2014kla}
A.~Ramos, {\it {The gradient flow running coupling with twisted boundary
  conditions}},  {\em JHEP} {\bf 11} (2014) 101,
  [\href{http://arxiv.org/abs/1409.1445}{{\tt arXiv:1409.1445}}].

\bibitem{Perez:2014xga}
M.~Garc\'{\i}a~P\'erez, A.~Gonz\'alez-Arroyo, L.~Keegan, and M.~Okawa, {\it
  {TEK twisted gradient flow running coupling}},  {\em PoS} {\bf LATTICE2014}
  (2014) 300, [\href{http://arxiv.org/abs/1411.0258}{{\tt arXiv:1411.0258}}].

\bibitem{GonzalezArroyo:1997uj}
A.~Gonz\'alez-Arroyo, {\it {Yang-Mills fields on the 4-dimensional torus. Part
  I: Classical Theory}},  {\em World Scientific. Proceedings of the Pe\~niscola
  1997 advanced school on non-perturbative quantum field physics, Singapore}
  (1998) [\href{http://arxiv.org/abs/hep-th/9807108}{{\tt hep-th/9807108}}].

\bibitem{Luscher:2011bx}
M.~Luscher and P.~Weisz, {\it {Perturbative analysis of the gradient flow in
  non-abelian gauge theories}},  {\em JHEP} {\bf 02} (2011) 051,
  [\href{http://arxiv.org/abs/1101.0963}{{\tt arXiv:1101.0963}}].

\bibitem{Fodor:2012td}
Z.~Fodor, K.~Holland, J.~Kuti, D.~Nogradi, and C.~H. Wong, {\it {The Yang-Mills
  gradient flow in finite volume}},  {\em JHEP} {\bf 11} (2012) 007,
  [\href{http://arxiv.org/abs/1208.1051}{{\tt arXiv:1208.1051}}].

\bibitem{Fritzsch:2013je}
P.~Fritzsch and A.~Ramos, {\it {The gradient flow coupling in the Schr\"odinger
  Functional}},  {\em JHEP} {\bf 1310} (2013) 008,
  [\href{http://arxiv.org/abs/1301.4388}{{\tt arXiv:1301.4388}}].

\bibitem{Ramos:2014kka}
A.~Ramos and S.~Sint, {\it {On ${\cal O}(a^2)$ effects in gradient flow
  observables}},  {\em PoS} {\bf LATTICE2014} (2014) 107,
  [\href{http://arxiv.org/abs/1411.6706}{{\tt arXiv:1411.6706}}].

\bibitem{sintramos2}
A.~Ramos, {\it {Wilson flow and renormalization}},  {\em PoS} {\bf LATTICE2014}
  (2014) 079.

\bibitem{Snippe:1996bk}
J.~R. Snippe, {\it {Square Symanzik action to one-loop order}},  {\em Phys.
  Lett.} {\bf B389} (1996) 119,
  [\href{http://arxiv.org/abs/hep-lat/9608146}{{\tt hep-lat/9608146}}].

\bibitem{Snippe:1997ru}
J.~R. Snippe, {\it {Computation of the one-loop Symanzik coefficients for the
  square action}},  {\em Nucl. Phys.} {\bf B498} (1997) 347,
  [\href{http://arxiv.org/abs/hep-lat/9701002}{{\tt hep-lat/9701002}}].

\bibitem{Fodor:2014cpa}
Z.~Fodor, K.~Holland, J.~Kuti, S.~Mondal, D.~Nogradi, et~al., {\it {The lattice
  gradient flow at tree-level and its improvement}},  {\em JHEP} {\bf 1409}
  (2014) 018, [\href{http://arxiv.org/abs/1406.0827}{{\tt arXiv:1406.0827}}].

\end{thebibliography}\endgroup

\end{document}